\documentclass[manuscript]{aastex}
\usepackage{natbib}
\usepackage{bm}
\usepackage{enumerate}
\usepackage{float}
\newcommand{\brac}[1]{\langle #1 \rangle}
\newcommand{\pd}{\partial}

\newcommand{\mean}[1]{\overline{#1}}

\def\Ra{\mbox{\rm Ra}}
\def\Ma{\mbox{\rm Ma}}

\def\Ro{\mbox{\rm Ro}}

\def\Pra{\mbox{\rm Pr}}

\def\cp{c_{\rm p}}

\def\Rs{R_{\odot}}
\def\cs{c_{\rm s}}

\def\urms{u_{\rm rms}}

\usepackage{graphicx}
\usepackage{graphics}
\graphicspath{{fig_lr/}{png/}}

\slugcomment{Not to appear in Nonlearned J., 45.}

\shorttitle{Global simulation of convection}
\shortauthors{Guerrero et al.}
\begin{document}

\title{Differential rotation in solar-like stars from global simulations}


\author{G. Guerrero}
\affil{Solar Physics, HEPL, Stanford University, 
452 Lomita mall, Stanford, CA, 94305-4085}
\email{gag@stanford.edu}

\author{P.~K. Smolarkiewicz}
\affil{European Centre for Medium-Range Weather Forecasts, Reading RG2 9AX, UK}
\email{smolar@ecmwf.int}

\author{A.G. Kosovichev}
\affil{{Solar Physics, HEPL, Stanford University, 
452 Lomita mall, Stanford, CA, 94305-4085}}
\affil{Big Bear Solar Observatory, NJIT, 40386 North Shore Lane, 
Big Bear City, CA 92314-9672, U.S.A.}
\email{sasha@sun.stanford.edu}

\and
\author{N. N. Mansour}
\affil{NASA, Ames Research Center, Moffett Field, Mountain View, CA 94040, USA}
\email{Nagi.N.Mansour@nasa.gov}

\begin{abstract}
To explore the physics of large-scale flows in solar-like stars, we
perform 3D anelastic simulations of rotating convection for global
models with stratification resembling the solar interior. The
numerical method is based on an implicit large-eddy simulation
approach designed to capture effects from non-resolved small scales.
We obtain two regimes of differential rotation, with equatorial
zonal flows accelerated either in the direction of rotation
(solar-like) or in the opposite direction (anti-solar). While the
models with the solar-like differential rotation tend to produce
multiple cells of meridional circulation, the models with anti-solar
differential rotation result in only one or two meridional cells.
Our simulations indicate that the rotation and large-scale flow
patterns critically depend on the ratio between buoyancy and Coriolis
forces. By including a subadiabatic layer at the bottom of the domain,
corresponding to the stratification of a radiative zone, we reproduce
a layer of strong radial shear similar to the solar tachocline.
Similarly, enhanced superadiabaticity at the top results in a
near-surface shear layer located mainly at lower latitudes.
The models reveal a latitudinal entropy gradient localized at the base 
of the convection zone and in the stable region, which however does 
not propagate across the convection zone.
In consequence, baroclinicity effects remain small and the rotation
iso-contours align in cylinders along the rotation axis. Our results
confirm the alignment of large convective cells along the rotation
axis in the deep convection zone, and suggest that such
"banana-cell" pattern can be hidden beneath the supergranulation
layer.
\end{abstract}





\section{Introduction}

Stellar magnetic activity is thought to be a result
of inductive effects of large-scale shearing flows as well as
complicated collective effects of small-scale turbulent motions in 
convective envelopes.  Cyclic behaviour has been detected in 
several late-type stars in the emission cores
of the Calcium H and K line profiles, with periods between 7 and 22 years 
\citep{BV85,Hall08}.  The best known and studied example is, however,
the 11-years sunspot cycle, where observed activity is 
caused by yet poorly understood 
dynamo processes in the convection zone occupying the upper 30\% of
the solar radius. A similar mechanism is expected to operate
in other solar-like stars.

Several different dynamo models have been proposed
over the years in order to describe the solar cycle using 
a MHD mean-field \citep{SKR66} formalism \citep[see][for 
reviews on the topic]{BS05,Ch10}. However, since the dynamics of 
the magnetic field in the solar interior
remains hidden to the observations, there are strong model ambiguities.

The large-scale plasma flows in the solar
interior, such as differential rotation and meridional circulation have
been probed thanks to helioseismology. 
We know that the Sun rotates differentially in latitude through 
the whole convection zone with a profile that forms conical 
iso-rotation contours \citep{schou+etal_98}.  
Below the bottom of
the convection zone the rotation becomes rigid creating a thin
radial shear layer \citep{K96} called the tachocline \citep{SZ92}. 
Besides, at the upper
50Mm of the convection zone the rotation rate decreases 
apparently at all latitudes forming another thin layer of 
negative radial shear so-called near-surface shear layer \citep{Tetal96,CT02}.

The meridional circulation has been observed at the solar surface
through different techniques \citep[e.g.,][]{GDSB97,ZK04,UL10, HR10, GR08}.
This flow is directed poleward with amplitudes
$\sim 20$ m/s peaking at mid latitudes.  By tracking supergranules 
\cite{Ha12} was recently able to infer a return flow at $~70$ Mm below 
the photosphere. Besides, helioseismic inversions by \cite{ZBKD12,ZBKD13} 
revealed that the return flow is located in the middle of the convection 
zone and suggested the existence of a second circulation cell located deeper.

More recently asteroseismology opened the door to the 
exciting possibility of measuring the large scale flows in different
classes of stars. Of special interest are stars at the same stage
of evolution and structurally similar to the Sun.  The study of
these stars could provide a better understanding of the stellar 
dynamo mechanism and the generation of cosmic magnetic 
fields in general.

The observational results from helioseismology have
imposed some constrains to the mean-field dynamo models but are still
insufficient to  provide a complete scenario. Thus, global 
numerical simulations seem to be the most promising approach towards
the understanding of the solar dynamics and dynamo, filling the gap 
left by observations.  
Steady dynamo solutions have been found in numerical models at the solar
rotation rate. These models showed that the magnetic field in the convection
zone is organized in the form of toroidal wreaths \citep{BBBMT10}.
Oscillatory dynamo solutions have also
been found in simulations of stars rotating at $3$ or more times faster than 
the Sun \citep{KMB12,NBBMT13}. However, none of the current
numerical models has been able to reproduce sufficiently well the solar 
differential rotation and meridional circulation, in particular, the 
tachocline  and the subsurface shear layer, and still are not able to  
explain the solar activity cycles.

In spite of the fast developments of computer capabilities,
all global simulations are far from realistically modeling the
Sun or solar-like stars. The problem arises from the 
low values of the dissipative coefficients in the stellar interior
which impose a large separation among the scales of energy
injection and diffusion. 
Most of the simulations fall in the 
group of the so-called Large-Eddy Simulation (LES) models. 
These models resolve explicitly the evolution of large-scale
motions while the contribution
of unresolved, sub-grid scale (SGS), motions is represented by
{\it ad-hoc} turbulence models \citep[e.g.][]{Sma63}.
For many years
the small-scale contribution used in solar global simulations was
the most simple, e.g., in the form of enhanced eddy viscosity
\citep[e.g.,][]{Gi76,BMT04}. Recently, more sophisticated SGS
models, like the dynamic Smagorinsky viscosity model \citep{GPMC91}, 
have been implemented leading to a more turbulent regime in the global 
models \citep{NBBMT13}.

An interesting alternative to such explicit SGS modeling consists in
developing numerical schemes which accurately 
model the inviscid evolution of the large scales whilst the SGS
contribution is intrinsically captured by numerical dissipation.
This class of modeling is called Implicit Large Eddy Simulations
(ILES), and is implemented in the EULAG code \citep{S06,PSW08},
used primarily in atmospheric and climate research.

Recently,  oscillatory dynamo solutions were studied in global 
simulations performed with the EULAG code for the case of
the solar rotation \citep{GCS10}. Although the period
of the simulated activity cycles is of $\sim40$ years
(instead of 11), and also 
there is no clear latitudinal migration of the toroidal
field towards the equator, these results are a promising step forward in the
understanding of the solar dynamo. These simulations
with EULAG were mostly focused on the dynamo rather than on
the differential rotation problem. The present work aims to
explore the hydrodynamical case of rotating convection with
the EULAG code, giving attention to the physics behind the 
development of large-scale flows: meridional circulation and
differential rotation.

The main challenge is to reproduce the solar differential 
rotation profile inferred by helioseismology with the 
characteristics described above. Starting with the seminal
work by \cite{Gi76} this problem has been studied numerically 
during the last few decades without obtaining a completely 
satisfactory result. This is a complex problem, and its 
solution must fulfill several observed properties: 
\begin{enumerate}[i]
\item reproduce the latitudinal differential 
rotation with a difference in the rotation rate between 
the equator and $60^{\circ}$
latitude of $\sim 90$nHz;
\item obtain iso-rotation contours of $\Omega(r,\theta)$ 
aligned along conical lines in the bulk of the convection zone;
\item reproduce the sharp transition region between the
differentially rotating convection zone and the solid
body rotating radiative zone, i.e., the tachocline,
\item  reproduce a sharp decrease in the solar rotation rate in
the upper 50Mm of the solar convection zone, in other words,
a near-surface rotational shear layer with a negative radial
gradient of the angular velocity.
\end{enumerate}

The item i) has been explored since \cite{Gi76}, and is understood
as a combined effect of the buoyancy and Coriolis forces. 
Rotation leads to correlations
between the three components of convective motions, reflected in
anisotropy of
the Reynolds stress tensor, which determines the direction of the
angular momentum flux.  In global simulations with the 
anelastic code ASH,  \cite{BT02} 
scanned the parameter space, particularly in terms of the Rayleigh 
($\Ra$) and Prandtl ($\Pra$) numbers, to determine  
a trend at which the 
latitudinal gradient of differential rotation could be sustained at 
higher levels of turbulence. 

Item ii) is perhaps the most cumbersome of
all the items above. Most of the global simulations of rotating
convection obtain the iso-lines of angular velocity aligned along the 
rotation axis, following the so-called Taylor-Prouman state, which can 
be explained 
from the $\phi$ component of the vorticity equation:

\begin{equation}
  \frac{\partial \omega_{\phi}}{\partial t} =
  r \sin\theta \frac{\partial \Omega^2}{\partial z} -
  \frac{g}{r C_p} \frac{\partial s}{\partial \theta}\;,
\label{eq.tp}
\end{equation}
where $\omega_{\phi}$ is the azimuthal component of the vorticity, 
$\Omega$ is the angular velocity, $g$ is the gravity acceleration,
$s$ is the entropy and $c_p$ is the specific heat at constant pressure.
The derivative along the cylindrical radius is defined as
$\partial_z=\cos\theta \partial_r - r^{-1} \sin\theta \partial_{\theta}$.  

In the steady state ($\partial \omega_{\phi}/\partial t=0$), 
the cylindrical iso-rotation could be attributed to insufficient 
baroclinicity in the system, i.e., if entropy does not exhibit 
a pronounced latitudinal gradient then 
${\partial \Omega^2}/{\partial z}=0$ . 
In hydrodynamic mean-field models, the Taylor-Proudman state 
is broken by considering anisotropic heat conductivity 
which could result from rotational 
influence on the turbulent eddies \citep{KR95}.  Similarly, \cite{R05}
studied how a subadiabatic stratification beneath the convection zone
can produce a gradient in the entropy perturbations
that could be transported into the convection zone.  
\cite{MBT06} implemented the Rempel's idea in
global convection simulations by imposing a latitudinal
gradient of entropy as a bottom boundary condition.
They found with such model that the iso-contours of angular
velocity form conical lines such as determined by helioseismology.

As for the item  iii), the tachocline has been included
in dynamo simulations by imposing a forcing term in the
momentum equation \citep{BMBT06}.  Hydrodynamic simulations
by \cite{KMGBC11} considered a stable radiative zone at 
the bottom of the domain. However, due to the large values of the
viscosity the angular momentum was transported into
the sub-adiabatic zone causing the spreading of the tachocline.  
Simulations with the EULAG code \citep[][and the ones presented 
in this paper]{GCS10,RCGS11,GSKM13} include a strongly 
sub-adiabatic radiative zone, and
have been able to  reproduce the rotational shear
layer in agreement with the solar observations. The key 
elements in the tachocline modeling  
are the stably stratified layer at the bottom of the domain
and small values of the numerical viscosity, preventing
the transport of angular momentum into the radiative zone.
Finally, for item iv) \cite{dRGT02} studied convection in
a thin shell convering only a fraction of the upper 
convection zone. With the latitudinal solar rotation 
imposed as a boundary condition at the bottom of the 
domain, they found a decrease in the rotation rate 
with height at each latitude resembling the observations.
However, the formation of this layer has not
been explored in global models including the whole 
convection zone. 

It is also a challenge to reproduce the poleward meridional 
circulation  observed at the solar
surface. This is more difficult to achieve since
global numerical models are not able to resolve the large-scale 
dynamics in the bulk of the convection zone together with the 
small-scale motions, such us granulation and supergranulation, 
observed in the uppermost layers.  
Hydrodynamical mean-field models are
able to reproduce the Sun's poleward meridional velocity 
\citep{R05,KO12}, and obtain one meridional
cell per hemisphere. In fact, \cite{KO12} have obtained in their
models a single meridional circulation cell for different late-type 
stars, independenlty of the stratification or the rotation rate.
This result is in disagreement with most of the global numerical
simulations which normally show that the meridional circulation has
several cells per hemisphere.
   
Furthermore, recent studies of the amplitude of  
convective velocities at the solar interior have pointed out the
discrepancy between the turbulent velocities obtained by 
helioseismology observations \citep{HDS12}, theoretical 
estimations \citep{MFRT12} and numerical models 
\citep[e.g.,][]{MBdRT08}.
This has emphasised the need of a better understanding of the 
multiscale turbulent velocities in the solar convection zone.  Both 
issues, the meridional circulation and turbulent velocities 
deserve further investigation.

Having a solar model as our reference, in this paper we investigate 
the items  (i) to (iv) for the
Sun and also study the generation of differential rotation 
and meridional flows in solar-like stars. Specifically,
we study turbulent convection in spherical shells, the
stratification of which resembles the structure of the solar 
interior.  Considering simulations for different rotation 
rates we first explore the 
phenomena of angular momentum transport and its relationship
to different configurations of mean flows
within the convective layer. As a test for the implicit SGS
method employed in this paper,  we next explore the
convergence of the numerical scheme with increasing 
resolution.
Finally, considering angular momentum transport arguments 
we explore the possibility of forming a near-surface rotational
shear layer such as observed in the upper layers of the solar
convection zone. 

This paper is organized as follows: in  \S\ref{sec.model} we
describe the numerical model, in \S\ref{sec.results}
we present our results, including the properties of  
convective flows and the resulting differential rotation and
meridional circulation. We also describe the angular momentum 
mixing in terms of the 
Reynolds stresses. Finally we discuss our attempt to reproduce
the near-surface shear layer. We conclude and discuss our 
results in \S\ref{sec.conc}.

\section{Model}
\label{sec.model}

Similarly to the previous dynamo simulations with the EULAG code 
\citep{GCS10,RCGS11}, our model covers a full spherical shell,
i.e., $0\le \theta \le \pi$, and $0 \le \phi \le 2\pi$ in latitude
and longitude, respectively. In radius our domain spans from  
$0.61 \Rs$ to $0.96\Rs$ (although in section \ref{sec.nssl} we 
extend our domain up to $r=0.985\Rs$). 

We solve the anelastic set of hydrodynamic equations following
the formulation of \cite{LH82}:
\begin{equation}
          {\bm \nabla}\cdot(\rho_s\bm u)=0, \label{equ:cont}
\end{equation}
\begin{equation}
  \frac{D \bm u}{Dt}+ 2{\bm \Omega} \times {\bm u} =  
  -{\bm \nabla}\left(\frac{p'}{\rho_s}\right) + {\bf g}\frac{\Theta'}
  {\Theta_s} \;, \label{equ:mom} 
\end{equation}
\begin{equation}
 \frac{D \Theta'}{Dt} = -{\bm u}\cdot {\bm \nabla}\Theta_e + \frac{1}{\rho_s}
                        {\cal H}(\Theta')-\alpha\Theta'\;,
 \label{equ:en} 
\end{equation}
\noindent
where $D/Dt = \pd/\pd t + \bm{u} \cdot {\bm \nabla}$ is the total
time derivative, ${\bm u}$, is the velocity field in a rotating 
frame with constant angular velocity ${\bm \Omega}$,
$p'$ and $\Theta'$ are the
pressure and potential temperature fluctuations with respect to an
ambient state (discussed below), respectively, 
$\cal{H}(\theta')$ represents the radiative and heat diffusion terms, 
$\rho_s$  and $\Theta_s$ are the density and potential temperature 
of the reference state chosen to be isentropic (i.e., $\Theta_s={\rm const}$)
and in hydrostatic equilibrium,  $g=GM/r^2$ is the gravity acceleration. 
The potential temperature, $\Theta$, is related to the specific
entropy through $s=c_p \ln\Theta+{\rm const}$. 
 
Finally, the term   $\alpha\Theta'$ represents the balancing 
action of the turbulent heat flux  responsible for 
maintaining the steady, axi-symmetric, solution of the stellar structure 
(see next section). It is characterised by the ambient state $\Theta_e$ 
(together with the corresponding pressure $p_e$). In the reported simulations, 
the $\alpha\Theta'$ term forces the system towards the ambient state on a 
time scale $\tau =\alpha^{-1}=1.55\times10^8$ s ($\sim 5$ years).
This timescale is much 
shorter than the timescale of radiative losses and heat diffusion, 
meaning that the turbulent heat flux is the primary driver of
convection in the system \citep[see also][]{CCS13}. After verifying
that the terms in $\cal{H}(\theta')$ do not have any effect on the
model solutions,  we did not included these terms in the simulations 
below. However, the choice of $\tau$ can affect the results since
the final potential temperature is the total of the 
ambient state and fluctuations. The longer (shorter) $\tau$, the larger
(smaller) the value of $\Theta'$. We have verified (in simulations not 
presented here) that models with shorter $\tau$ suppress convection and 
reproduce a profile close to $\Theta_e$. In models with longer $\tau$, 
on the other hand, the large fluctuations tend to homogenize the final 
potential temperature creating a flat, more adiabatic, profile.  
Consequently, the selected value of $\tau$ has been determined
empirically as a compromise between a weak forcing of the energy
equation and a reasonable relaxation time to a statistically steady
state of the system, that still leads to the desired results
while maintaining convection and the stellar structure over a
long observation time.

\subsection{Ambient state}
\begin{figure}[H]
\includegraphics[width=0.98\columnwidth]{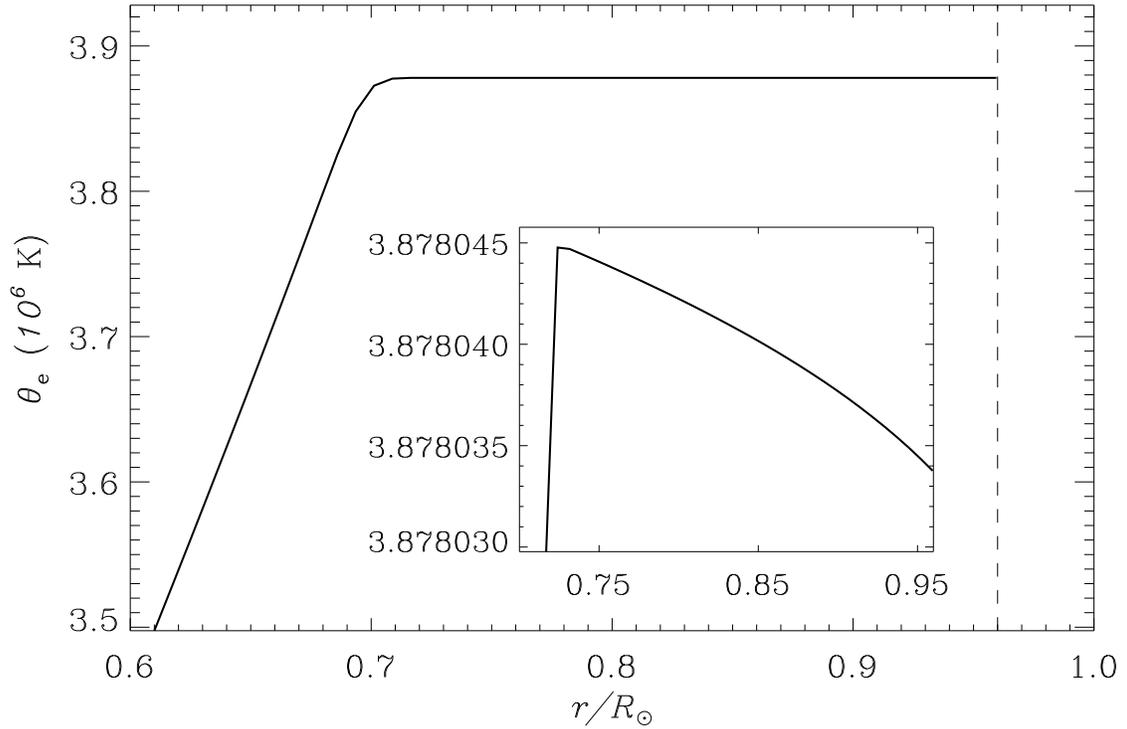}
\caption{Radial  profile of the potential temperature for the ambient 
state. The inset panel shows a zoom of the convection zone from $r=0.7\Rs$
to $r=0.96\Rs$.}
\label{fig.the}
\end{figure}

The ambient state in the simulations
is approximated by a polytropic model. The polytropic 
stratification for an ideal 
gas, $p=\rho R T$, is given by the hydrostatic equation:
\begin{equation}
\frac{dT}{dr} = \frac{g}{(1+m)R} \;,
\label{eq.poly}
\end{equation}
where $g(r)=g_b(r_b/r)^2$ is the gravity acceleration, $r_b$ is the 
radius of the bottom boundary, $g_b$ is the gravity acceleration at 
$r_b$, $R$ is
the gas constant, and $m$ is the polytropic index. The solution
of Eq. (\ref{eq.poly}) integrated from $r_b$ to $r$ and assuming
$m={\rm const}$ is:

\begin{eqnarray}
\label{eq.ps}
T_0 &=& T_b \left[1-\frac{r_b g_b r_b}{p_b (1+m)}
         \left(1- \frac{r_b}{r} \right) \right]\;,\\\nonumber
\rho_0 &=& \rho_b \left[1-\frac{\rho_b g_b r_b}{p_b(1+m)}
         \left(1 - \frac{r_b}{r}\right) \right]^m\;,\\\nonumber
p_0 &=& p_b \left[1-\frac{\rho_b g_b r_b}{p_b(1+m)}
         \left(1 - \frac{r_b}{r}\right) \right]^{m+1}\;,
\end{eqnarray}

here $T_b$, $\rho_b$ and $p_b$ are the values of 
the temperature, density and pressure at the bottom of the domain, 
$r_b$. These equations are solved
recursively from bottom to top for a polytropic index that varies
with radius as follows:
\begin{equation}
m(r)=m_r +  \Delta m \frac{1}{2}\left[1+ {\rm erf} \left( \frac{r-r_{tac}}
     {w_t} \right) \right] \;
\label{eq.pind}
\end{equation}

with $\Delta m = m_{cz}-m_{r}$, $m_{r} = 2.5$ and 
$m_{cz}=1.49995$. The potential temperature associated with this
stratification is computed through:
\begin{equation}
\Theta_e = T_0 \left( \frac{\rho_b T_b}{\rho_0 T_0} \right)^{1-1/\gamma},
\label{eq.the}
\end{equation}
where $\gamma=5/3$ defines the adiabatic exponent of an ideal gas. 
This  setup defines a strongly sub-adiabatic radiative layer   
(stable to convection) below  $r_{tac}=0.7\Rs$, followed by a slightly 
super-adiabatic, convectively unstable layer. An ${\rm erf}$ function
couples the two shells with a width of transition $w_t=0.01\Rs$
(see Fig. \ref{fig.the}).

\subsection{Numerical method and boundary conditions}

Eqs. (\ref{equ:mom}) to (\ref{equ:en}) are solved numerically 
using EULAG code \citep{SMW01,PSW08}.  EULAG is an HD and MHD
code for incompressible or anelastic fluid dynamics developed 
originally for research of atmospheric processes and climate and 
generalized over years for a range of
geophysical and astrophysical problems. The time evolution is
calculated using a unique semi-implicit method based on a high-resolution
non-oscillatory forward-in-time (NFT) advection scheme MPDATA
({\it Multidimensional Positive Definite Advection Transport
Algorithm}); see \cite{S06} for a recent overview. 
This numerical scheme does not require explicit dissipative
terms in order to remain stable. 
\cite{MR02} have demonstrated analytically that the  
numerical viscosity 
in the MPDATA method is comparable to 
the sub-grid scale eddy viscosity used in different LES models.
Although this rationale has been done for the 
solution of the Burger's equation only, 
MPDATA method has been successfully compared with LES in different
contexts \citep{ES02,DXS03,MSW06}. Thus, the results of
MPDATA are often interpreted as Implicit 
LES (ILES) models \citep{GMR07}. 

The boundary conditions are specified as follows.
In the latitudinal direction, discrete differentiation extends
across the poles, while flipping sign of the longitudinal and
meridional components of differentiated vector fields.
In the radial direction we use stress
free, impermeable, boundary conditions for the velocity field,
and asume that the normal derivative of
$\Theta'$ is zero. 
The simulations are initialized applying a small-amplitude 
random perturbation
to the velocity field; the results presented in the next
section are computed once the simulations have reached
a statistically steady state. It is characterized for the random
fluctuations of the volume averaged fluid quantities around a
well defined mean.

\section{Results}
\label{sec.results}

The convective properties of the non-rotating solution of 
Eqs. (\ref{equ:cont}-\ref{equ:en})
exhibit convection cells roughly of the same 
spatial extent with broad and slow
upflows and sharp and fast downflows (see Fig. \ref{fig.ur1}a).
The horizontal scale of this motion depends on the resolution
of the model. 
Initially  we consider the same number of grid points as 
in \cite{GCS10}, $n_{\phi}=128$,  $n_{\theta}=64$,
$n_r=47$. Since the viscous terms are not explicitly
considered in the equations, the standard Reynolds or
Taylor numbers, cannot be computed.  However, following 
\cite{CA06}, we consider a modified Rayleigh number which 
does not depend on the dissipative terms, 
$\Ra^*=\frac{1}{c_p \Omega_0^2} g \frac{\partial s_e}{\partial r}$,
where $s_e$ is the specific entropy of the ambient state.
From the solution we compute the Mach number, $\Ma=\urms/\cs^*$,
where $\urms$ is the volume averaged RMS turbulent velocity and
$\cs^*=\sqrt{\gamma R T_s^*}$, $T_s^*$ is the temperature of the
background state at the middle of the unstable layer ($r=0.85\Rs$);
and the Rossby number, which compares
the relative importance of turbulent convection over rotation:
$\Ro=\urms/(2\Omega_0 D)$, where $D$ is a typical length scale,
given here by the thickness of the convective unstable layer,
$0.25\Rs$.  In addition we compute a parameter characterizing 
the latitudinal differential rotation; 
\begin{equation}
\chi_{\Omega}=(\Omega_{eq}-\Omega_{p})/\Omega_0 \,
\label{eq.chi}
\end{equation}

where $\Omega_{eq}=\mean{\Omega}(\Rs,90^{\circ} {\rm colatitude})$  
and $\Omega_{p}=
\mean{\Omega}(\Rs,30^{\circ}{\rm colatitude})$. Overline
represents azimuthal averaging. 
In the Sun $\chi_{\Omega}\simeq0.2$. 
For models with different Rossby numbers  different values of 
$\chi_{\Omega}$ are expected due to different velocity correlations.
In the next section we study these correlations and the  
development of large scale motions in terms of the 
balance of angular momentum fluxes.

\begin{table}
\begin{center}
\caption{Simulation parameters and results. The quantities
in the table are defined as follows: 
$\Ra^*=\frac{1}{\cp \Omega_0^2} g \frac{\partial s_e}{\partial r}$, 
$\urms$ is the volume average rms velocity (in m/s) in the unstable layer,
$\Ma=\urms/\cs^*$ is the Mach number, with $\cs^*=\sqrt{\gamma R T_s^*}\;|_{r=0.85\Rs}$
being the sound speed at the middle of the unstable layer,
$\Ro=\frac{\urms}{2\Omega_0 L}$ and $\chi_{\Omega} = (\Omega_{eq}-\Omega_{p})/\Omega_0$, where
$\Omega_{eq}=\mean{\Omega}(\Rs,90^{\circ})$, $\Omega_{p}=\mean{\Omega}(\Rs,30^{\circ})$
and $\Omega_0=2.59\times10^6$ Hz.  Models starting with the letter ~L have the lowest 
resolution $N=(128\times64\times47)$, models starting with ~M and ~H have $2N$ and $4N$
grid points resolution, respectively. Model ~N1 has $256\times128\times100$ grid points. 
\label{tbl.1}}
\vspace{0.5cm}
\begin{tabular}{cccccccc}
\tableline\tableline
Model & $\Omega_0$ & $\Ra^*$ & $\urms$ & $\Ma(10^{-4})$ & $\Ro$ & $\chi_{\Omega}$ \\
\tableline
L0 &--                  &--   &39.6  &2.67   &--     &--\\
L1 &$\Omega_{\odot}/4$  &16.0 &36.2  &2.44   &0.160 &-0.39\\
L2 &$\Omega_{\odot}/2$  &4.00 &34.0  &2.29   &0.075 &-0.27 \\ 
L3 &$4\Omega_{\odot}/7$ &3.05 &34.8  &2.34   &0.067 &-0.29 \\
L4 &$4\Omega_{\odot}/6$ &2.24 &35.2  &2.37   &0.059 &0.03 \\
L5 &$4\Omega_{\odot}/5$ &1.56 &33.6  &2.26   &0.046 &0.07 \\
L6 &$\Omega_{\odot}$    &1.00 &28.3  &1.91   &0.031 &0.18 \\
L7 &$2\Omega_{\odot}$   &0.25 &28.1  &1.90   &0.016 &0.03 \\
\tableline
L8 &$\Omega_{\odot}$    &0.39 &13.7  &0.92   &0.015 & 0.12 \\
L9 &$\Omega_{\odot}$    &2.00 &47.0  &3.17   &0.052 &-0.22 \\
\tableline
M1 &$\Omega_{\odot}$    &1.02 &28.1  &1.90   &0.031 &0.16 \\ 
H0 &--                  &--   &36.8  &2.48   &--    &--   \\
H1 &$\Omega_{\odot}$    &1.10 &28.2  &1.90   &0.031 &0.09 \\
\tableline
N1 &$\Omega_{\odot}$    &1.78 &26.3  &1.77   &0.029 &0.087 \\
\tableline
\tableline
\end{tabular}
\end{center}
\end{table}

\subsection{Forces balance and angular momentum fluxes}
\label{sec.fbamf}
\begin{figure*}[h]
\includegraphics[width=0.48\columnwidth]{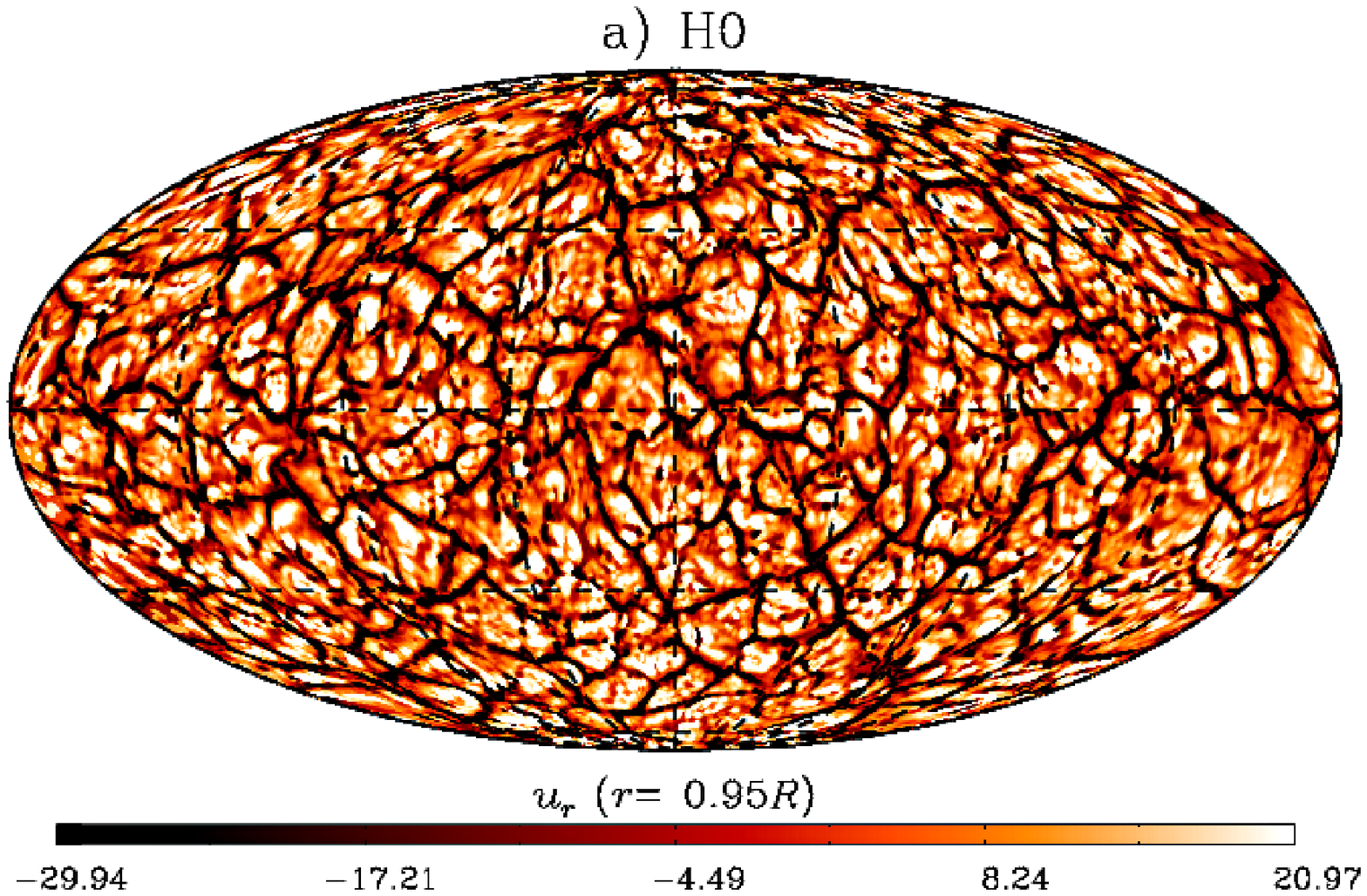}
\includegraphics[width=0.48\columnwidth]{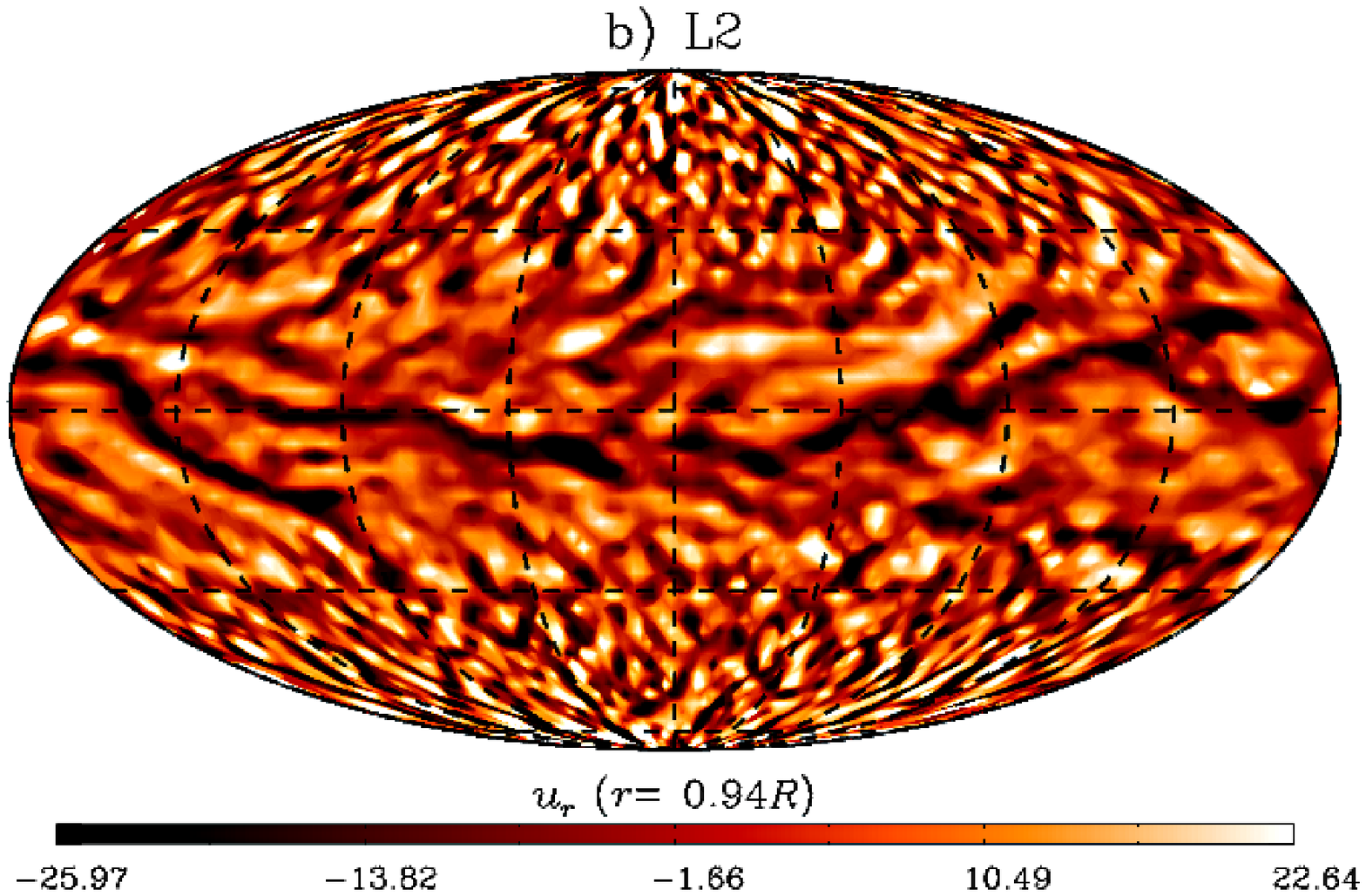}\\
\includegraphics[width=0.48\columnwidth]{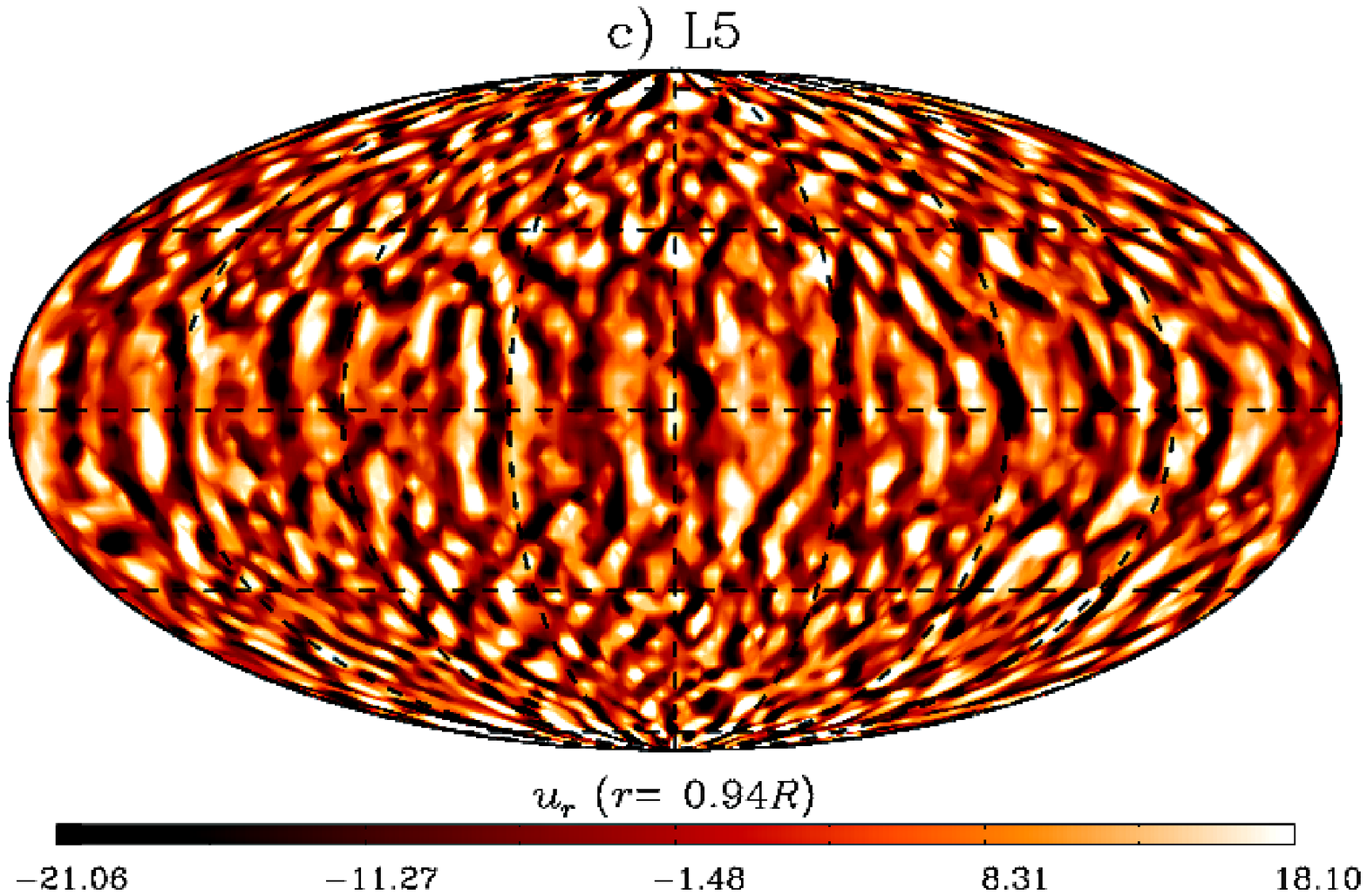}
\includegraphics[width=0.48\columnwidth]{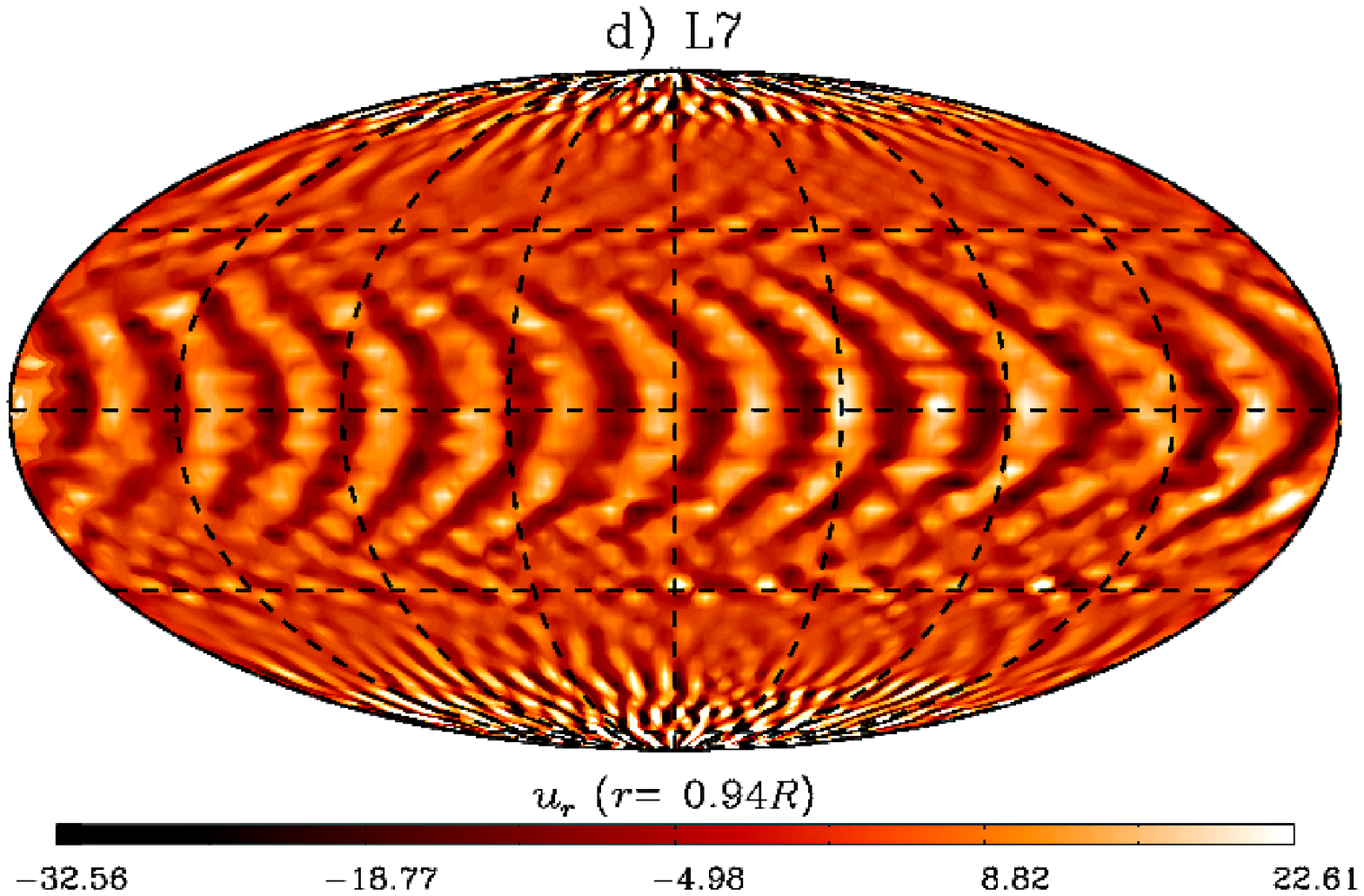}
\caption{Mollweide projection of the vertical velocity, $u_r$, 
at the subsurface layer, $r=0.95 \Rs$, for models ~H0 (without rotation), ~L2, 
~L5 and ~L7 (see Table \ref{tbl.1}). Yellow/dark color scale 
represent upward/downward motions.} 
\label{fig.ur1}
\end{figure*}

The development and sustenance of large-scale flows depend on 
correlations between the components of the turbulent 
velocities, the so-called Reynolds stresses.  The relative
importance of these quantities defines the redistribution
of angular momentum. Among these correlations, of particular 
importance are the vertical velocity and the
rotation rate of the system since they define the 
formation of fast rotating zonal flows either at higher
or lower latitudes. To understand this behaviour in our
first set of numerical experiments 
we change the rotation rate while keeping the 
ambient state constant (i.e., the same gradient of potential 
temperature, Eq. \ref{eq.the}).

\begin{figure*}[h]
\includegraphics[width=0.31\columnwidth]{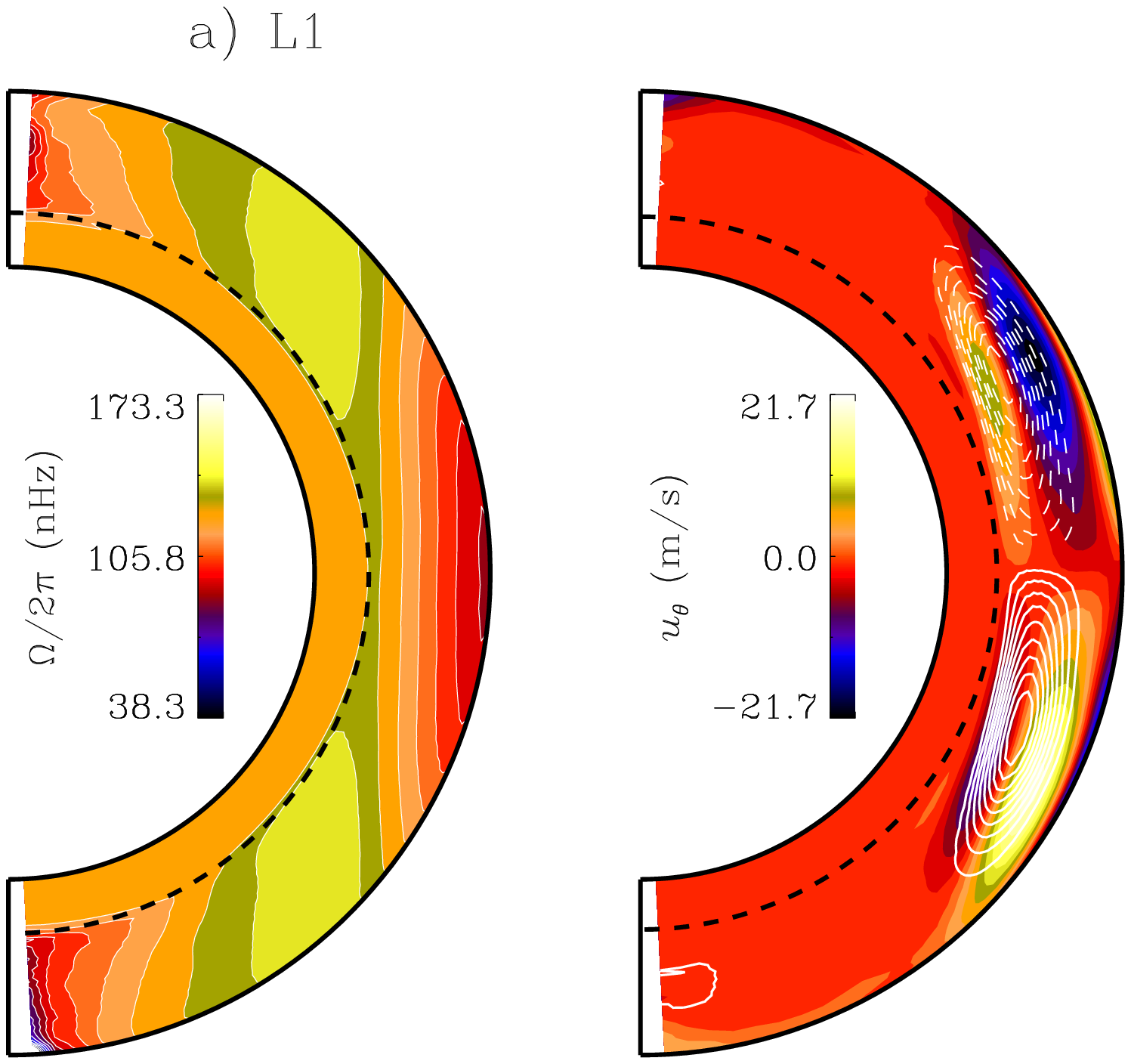}
\includegraphics[width=0.65\columnwidth]{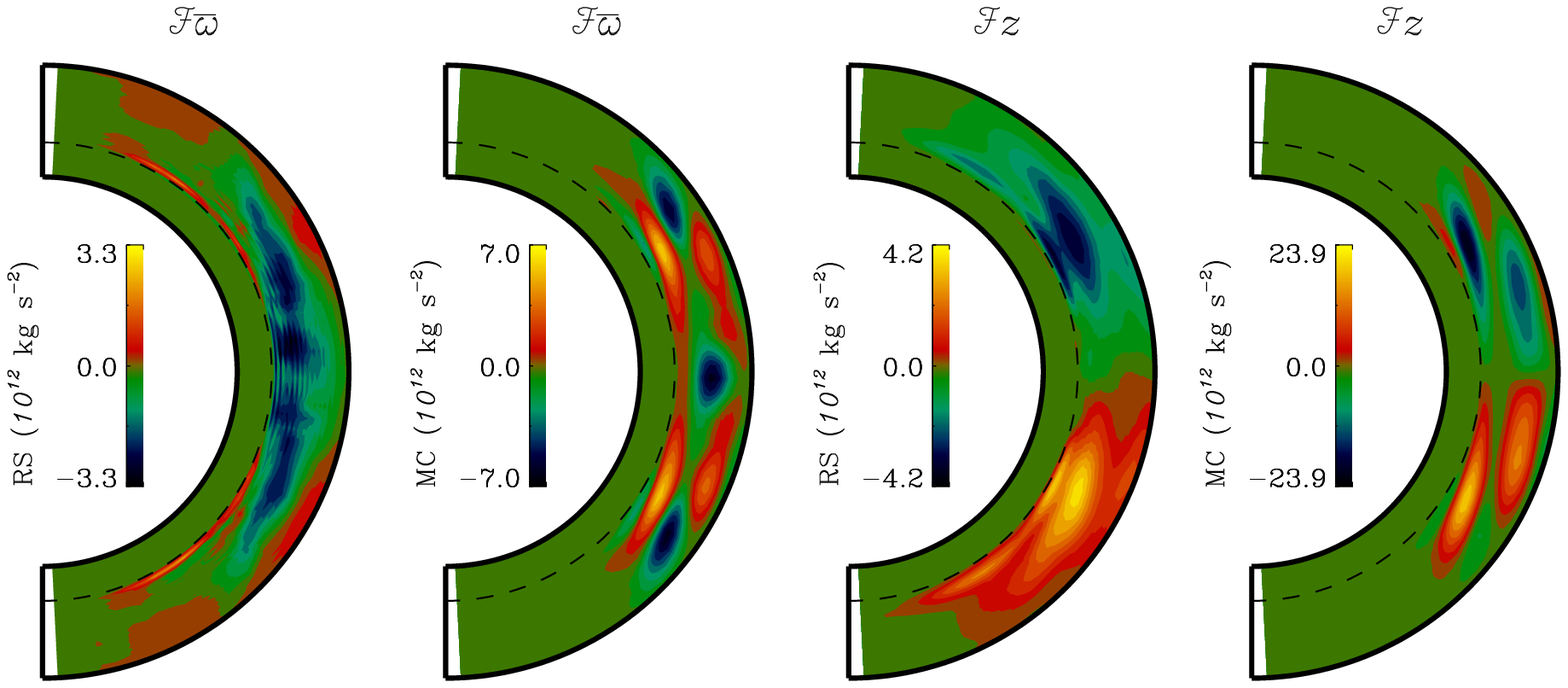}\\
\includegraphics[width=0.31\columnwidth]{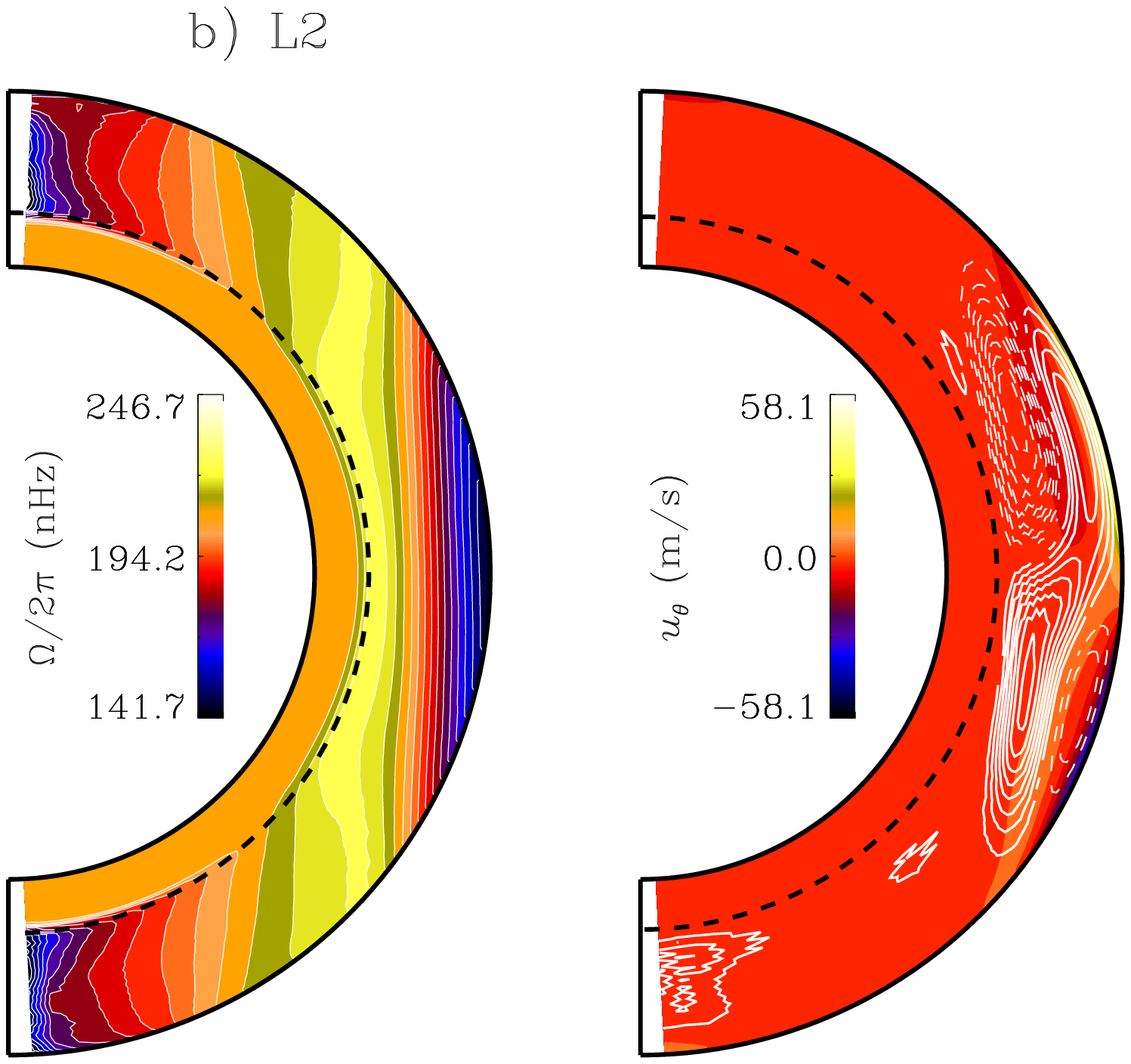}
\includegraphics[width=0.65\columnwidth]{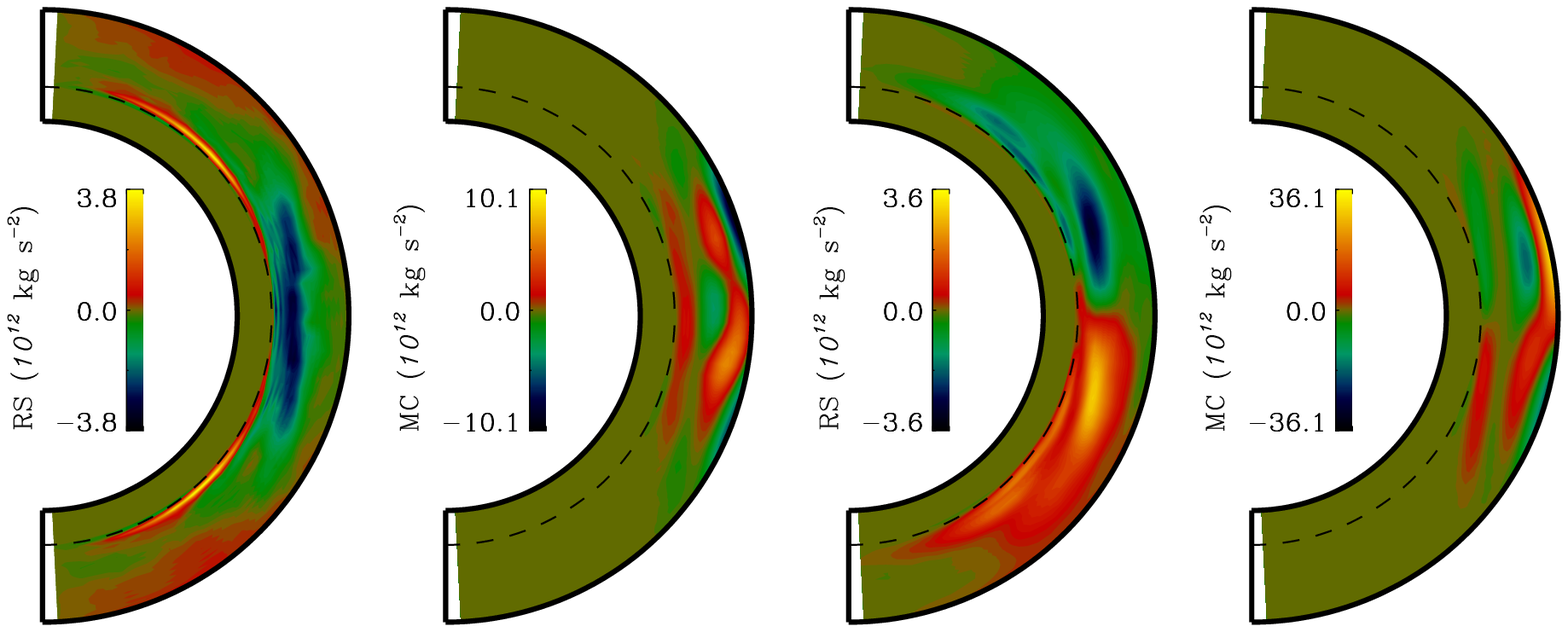}\\
\includegraphics[width=0.31\columnwidth]{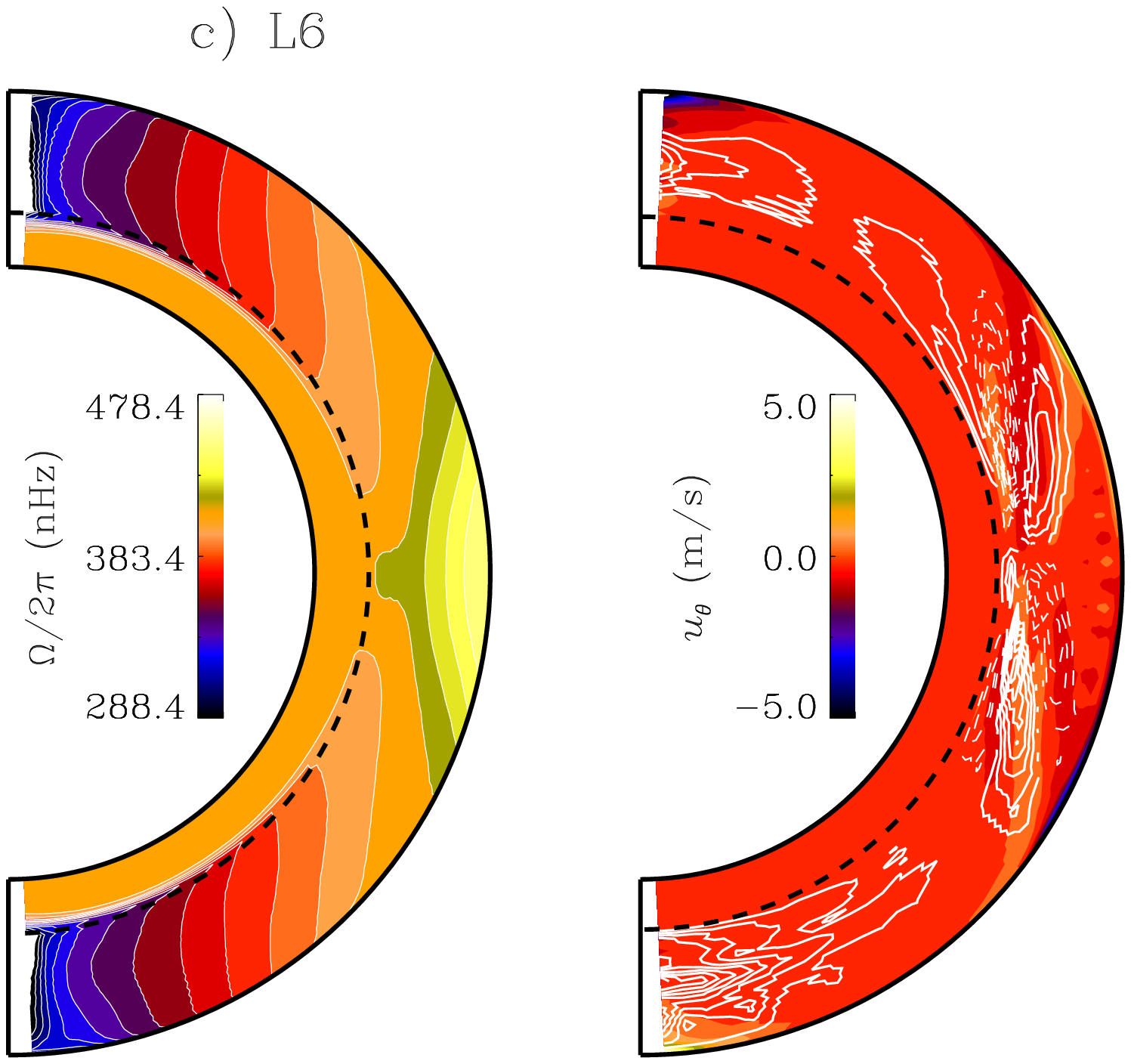}
\includegraphics[width=0.65\columnwidth]{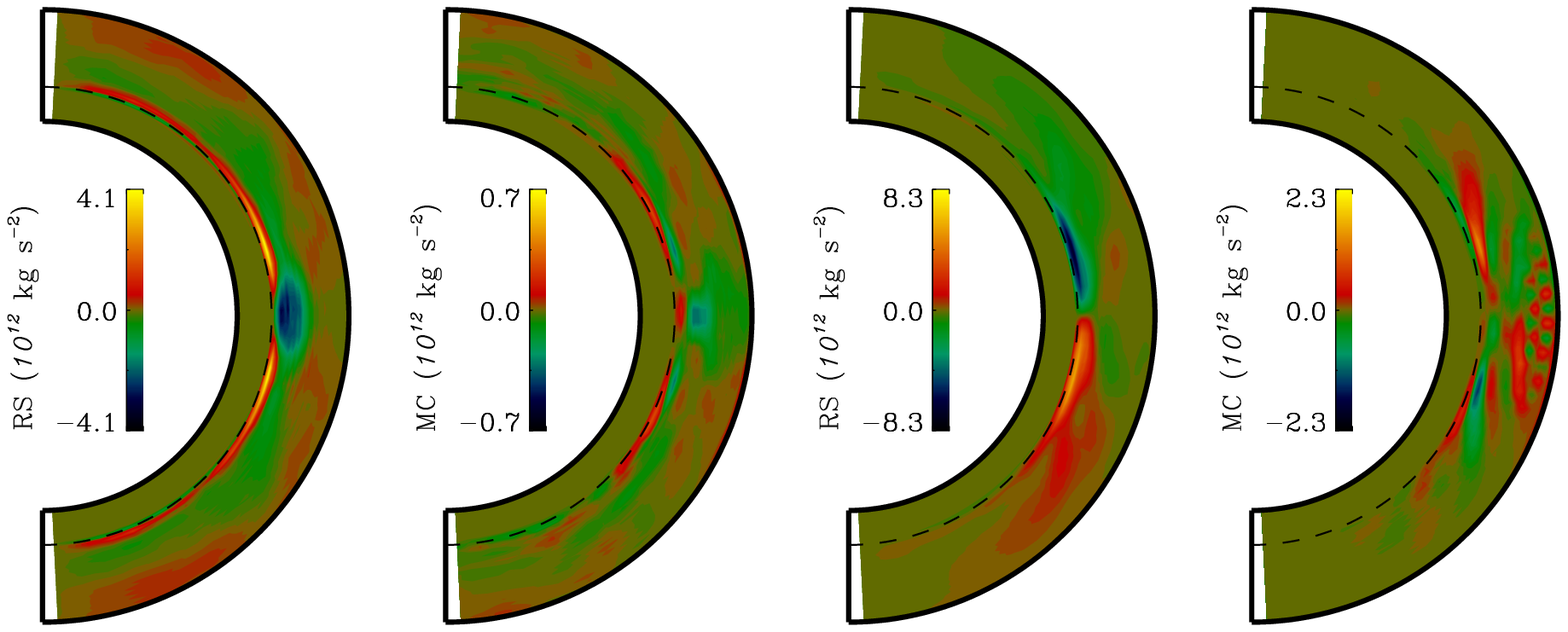}\\
\includegraphics[width=0.31\columnwidth]{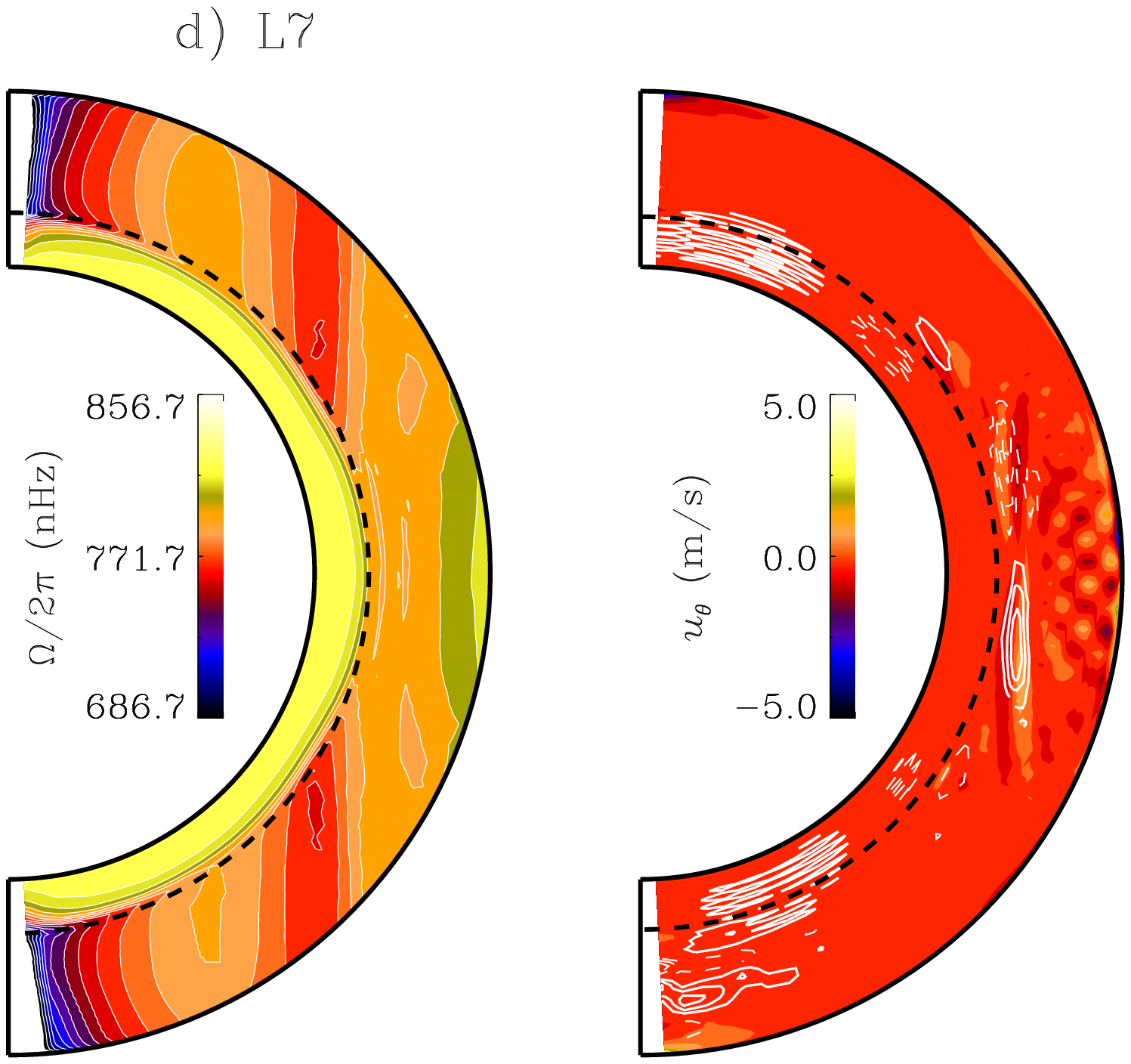}
\includegraphics[width=0.65\columnwidth]{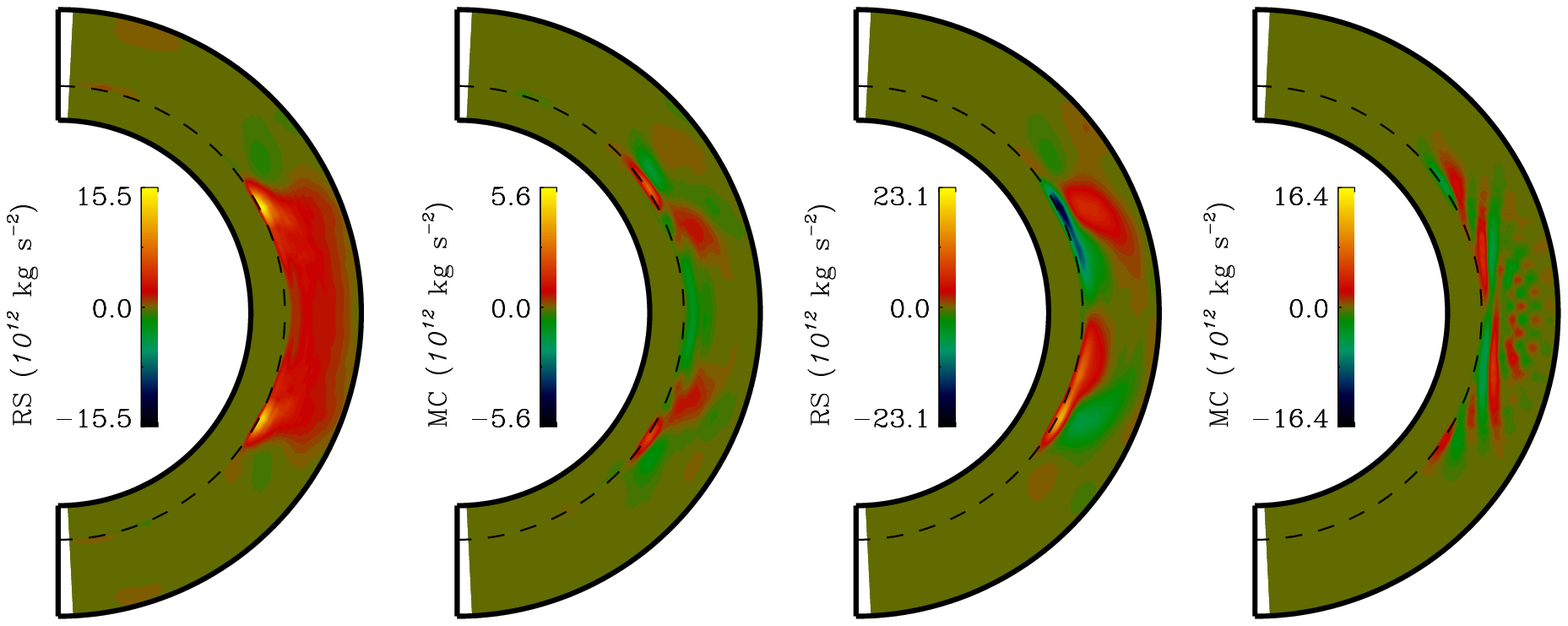}
\caption{Differential rotation, meridional circulation (panels 1 and
2 from left to right in each row) and angular momentum flux 
components of Eq. (\ref{eq.amf}):
${\cal F}_{\varpi}^{RS}$, ${\cal F}_{\varpi}^{MC}$, ${\cal F}_z^{RS}$ 
and ${\cal F}_z^{MC}$  (panels 3 to 6 from left to right),
for the simulations  ~L1, ~L2, ~L6 and ~L7 (top to bottom). All the 
profiles correspond to mean azimuthal value averaged over $\sim3$ 
years during the steady state phase of the simulation.
}
\label{fig.df1}
\end{figure*}

Models ~L1 to ~L7 in Table \ref{tbl.1}  summarize the 
parameters and results for this set of simulations. 
The structure of the convective pattern for some 
representative cases is shown in 
Fig. \ref{fig.ur1}b-d, where the radial component of the 
velocity at the top of the domain is presented using 
the Mollweide projection.  
The angular velocity, $\Omega/2\pi$, and the meridional 
circulation speed 
for some of these models are shown in the two leftmost panels
of Fig. \ref{fig.df1}.  We present the differential rotation
in nHz, in the meridional flow panel the color scale
corresponds to the latitudinal velocity in m/s, and the 
continuous (dashed) lines depict contours of the stream
function with clockwise (anti-clockwise) circulation. 
In the next four panels we show the fluxes of
angular momentum in the conservation equation (obtained by 
multiplying Eq. \ref{equ:mom} by the level arm $\varpi=r\sin\theta$):
\begin{equation}
\frac{\partial ({\rho}_s \mean{u}_{\phi})}{\partial t}
     =\frac{1}{\varpi}
      \nabla\cdot \left( {\rho}_s \varpi [
      \mean{u}_{\phi}\mean{\bm u}_{\rm m} + 
      \mean{u_{\phi}' {\bm u}'_{\rm m}} ]\right ),
\label{eq.amc1}
\end{equation}
where the over-line corresponds to the time and longitude 
average, $\mean{\bm u}_{\rm m}$ and ${\bm u}'_{\rm m}$ are the
mean and turbulent meridional ($r$ and $\theta$) components 
of the velocity, respectively. In steady state the left hand side 
of the equation vanishes remaining
\begin{equation}
      \nabla\cdot \left( {\rho}_s \varpi [
      \mean{u}_{\phi}\mean{\bm u}_{\rm m} + 
      \mean{u_{\phi}' {\bm u}'_{\rm m}} ]\right ) = 0.
\label{eq.amc1}
\end{equation}
The first term inside the divergence corresponds to the 
angular momentum flux due to 
meridional circulation and the second one is the flux due to 
small-scale correlations, the Reynolds stresses. 
Each of these terms has radial and latitudinal component,
\begin{eqnarray}
{\cal F}_r^{\rm MC}&=&\rho_s \varpi \mean{u}_{\phi} \mean{u}_r, \\\nonumber
{\cal F}_{\theta}^{\rm MC}&=&\rho_s \varpi \mean{u}_{\phi} \mean{u}_{\theta},\\\nonumber
{\cal F}_{r}^{\rm RS}&=&\rho_s \varpi \mean{{u}_{\phi}' {u}_{r}'}, \\\nonumber
{\cal F}_{\theta}^{\rm RS}&=&\rho_s \varpi \mean{{u}_{\phi}' {u}'_{\theta}}.
\end{eqnarray}
A more illustrative representation of these quantities is in
the cylindrical coordinates $\varpi$ and $z$ thus:
\begin{eqnarray}
\label{eq.amf}
{\cal F}_{\varpi}&=&{\cal F}_r \sin\theta + {\cal F}_{\theta} \cos\theta,\\\nonumber
{\cal F}_{z}&=&{\cal F}_r \sin\theta - {\cal F}_{\theta} \cos\theta.\\\nonumber
\end{eqnarray}
All these quantities were computed in the statistically steady 
state attained after a long time interval depending on the rotation rate,  
and averaged in time for 3.4 years. 
 
In the case of the slower rotation rate, $\Omega_0=\Omega_{\odot}/4$ 
(run ~L1),  the convective pattern
exhibits big convection cells which are slightly elongated in the 
direction opposite to the rotation. They span the range 
$\sim\pm45^{\circ}$ of latitude. 
As for the mean flows, the azimuthal motion 
is slower at the equator than at the higher latitudes 
(left panel of Fig. \ref{fig.df1}a).
The transport of angular momentum due to Reynold stress 
in the direction of $\varpi$ is  
negative at all latitudes (third panel of Fig. \ref{fig.df1}a), 
indicating that the rotation increases 
inwards. The most important contribution to the angular momentum
flux is, however, the meridional circulation
(rightmost panel of Fig. \ref{fig.df1}a).
Single large scale meridional flow cells
develops at each hemisphere. The circulation is counterclockwise 
(clockwise) in the northern (southern) hemisphere 
(second panel of Fig. \ref{fig.df1}a). This flow 
transports the angular
momentum from the lower to higher latitudes so that the 
zonal flow is accelerated in the direction of rotation at latitudes 
between $30^{\circ}$ and $80^{\circ}$. 

Increasing the rotation rate to $\Omega_0=\Omega_{\odot}/2$ 
(Run~L2) or 
$\Omega_0=0.57\Omega_{\odot}$ (Run~L3) we find that the convection cells are 
concentrated in a band around $\sim\pm30^{\circ}$ latitude. In this 
region the cells are mostly elongated in the longitudinal direction
(Fig. \ref{fig.ur1}b). 
Similarly to the previous model, the latitudinal differential rotation 
is anti-solar  (Fig. \ref{fig.df1}b). However,
the radial negative shear at the equator
starts not at the base of the convection zone, like in case ~L1,
but at $r\simeq0.73\Rs$. Between  $0.7\Rs < r < 0.73\Rs$ the plasma
rotates faster than the rotating frame. In this case the largest
contribution to the mixing of angular momentum comes again from 
the meridional circulation. However, this time two strong meridional
flow cells in radius are formed in each hemisphere.
In Run~L2 the latitudinal velocities 
are so high ($\sim40$ m/s) that the flow crosses from one 
hemisphere to another (second panel of Fig. \ref{fig.df1}b). 
Thus, the meridional flow is not fully 
symmetric across the equator even after a long temporal average. 

In runs ~L4 and ~L5 the contributions of meridional circulation and 
Reynold stresses to the flux of angular momentum
are similar. For these cases the plasma at
lower latitudes rotates roughly at the same rate as the rotating frame.
Above $\sim60^{\circ}$ latitude the zonal flow decelerates and has the 
minimal rotation rate at the poles. The meridional flow shows a 
multicellular pattern with three or more cells in radius per 
hemisphere, with latitudinal velocities 
of $\sim10$ m/s. 

Run ~L6 ($\Omega_0=\Omega_{\odot}$) shows a rotation profile
accelerated at the equator  (Fig. \ref{fig.df1}c). 
Similar to the Sun, the rotation
decreases monotonically towards the poles, and is in iso-rotation 
with the stable layer of the domain at middle latitudes. In this 
case the Reynolds stress components of the angular momentum 
flux dominate over the meridional circulation components. 
Although at lower latitudes, at the bottom
of the convection zone ${\cal F}^{RS}_{\varpi}$ is negative, it
changes sign above $r\sim0.85\Rs$. However, 
${\cal F}^{RS}_{z}$, which transports the angular momentum towards 
the equator is more important. Something similar happens in the 
fastest rotating case,
~L7 ($\Omega_0=2\Omega_{\odot}$,  Fig. \ref{fig.df1}d) 
where the Reynolds stress drive the differential
rotation. In this case, however, the transport along the rotation
axis, ${\cal F}^{RS}_{\varpi}$, is the most important flux component. 
The rotation is
faster at the equator and slower at the poles, but it exhibits jets
of slow rotation in a mid latitude range which corresponds to 
a cylinder tangent to the base of the convection zone. 
The term ${\cal F}^{MC}_{z}$ has larger values but varies quickly in 
radius and latitude.  Thus, there are no organized meridional 
circulation flows capable to effectively advect the angular momentum.
\begin{figure}[H]
\includegraphics[width=0.98\columnwidth]{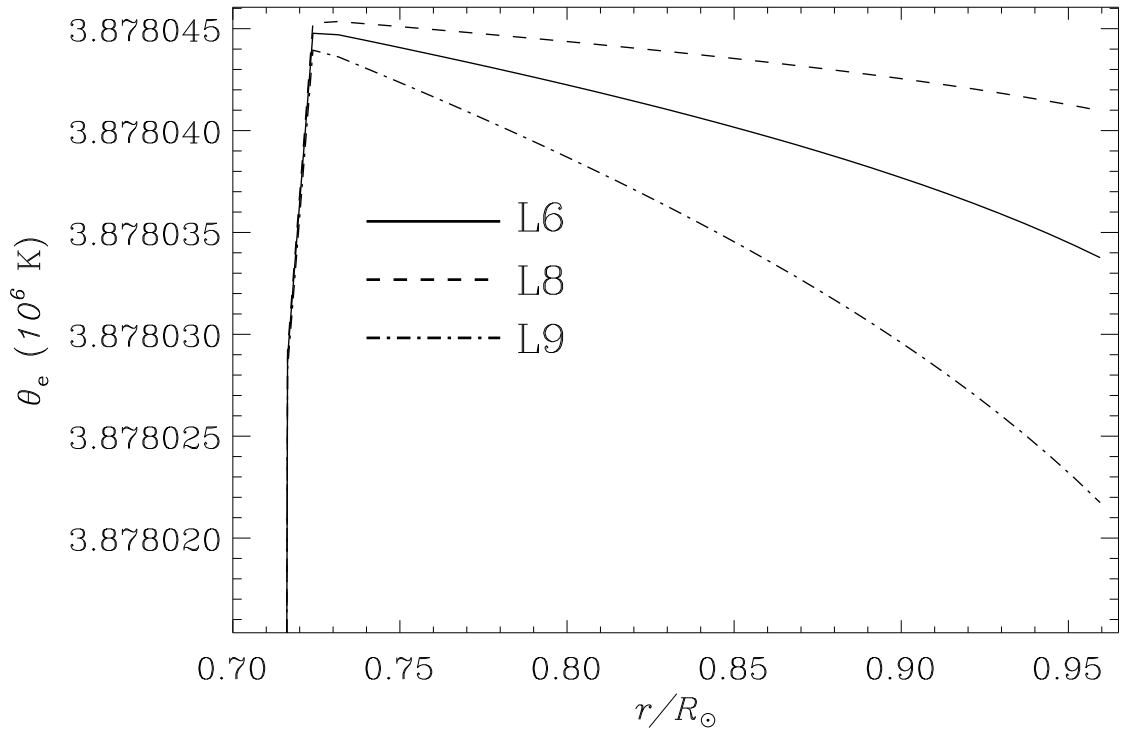}
\caption{Radial profiles of the $\Theta_e$ for models L6, L8 and L9 
as indicated in the legend.}
\label{fig.sevthe}
\end{figure}

Thus, in general, the differential rotation establishes as a 
competition
between the buoyant and Coriolis forces. When buoyancy dominates
then the differential rotation is of the anti-solar type, and 
when the Coriolis force rotation dominates
the rotation is solar-like. To demonstrate this point we have
performed the simulations labeled as ~L8 and L9 in Table \ref{tbl.1},
for which the rotation has the solar value, but the profile 
of $\Theta_e$ was modified 
to decrease and increase the buoyancy term by making the 
stratification less and more superadiabatic, respectively (see 
Fig. \ref{fig.sevthe}).  The model with less superadiabatic 
stratification, ~L8, develops a faster rotating equator as a 
consequence of the
dominant role of the Corioles force. The distribution of 
the angular momentum fluxes seems to be an intermediate case between
the models ~L6 and ~L7 (Fig. \ref{fig.df2}a). On the other hand, a more 
superadiabatic $\Theta_e$ results in antisolar rotation because
the buoyancy force dominates in this case. The results of this model
are compatible with the case ~L2 or ~L3 (Fig. \ref{fig.df2}b). 
Another way to modify the adiabaticity of the system is 
considering different relaxation times, $\tau$ 
($1/\alpha$ in Eq. \ref{equ:en}).
We have verified in simulations not presented here that models with
shorter $\tau$ develop small fluctuations of $\Theta'$, so that the
final profile of $\Theta$ is more superadiabatic.  Also, models with
longer $\tau$ develop large fluctuations that tend to homogenize the
potential temperature in the bulk of the convection zone, leading
to a less superadiabatic system.  Changes in $\tau$, however, are
not as important as directly modifying the profile of $\Theta_e$. 

\begin{figure}[H]
\includegraphics[width=0.32\columnwidth]{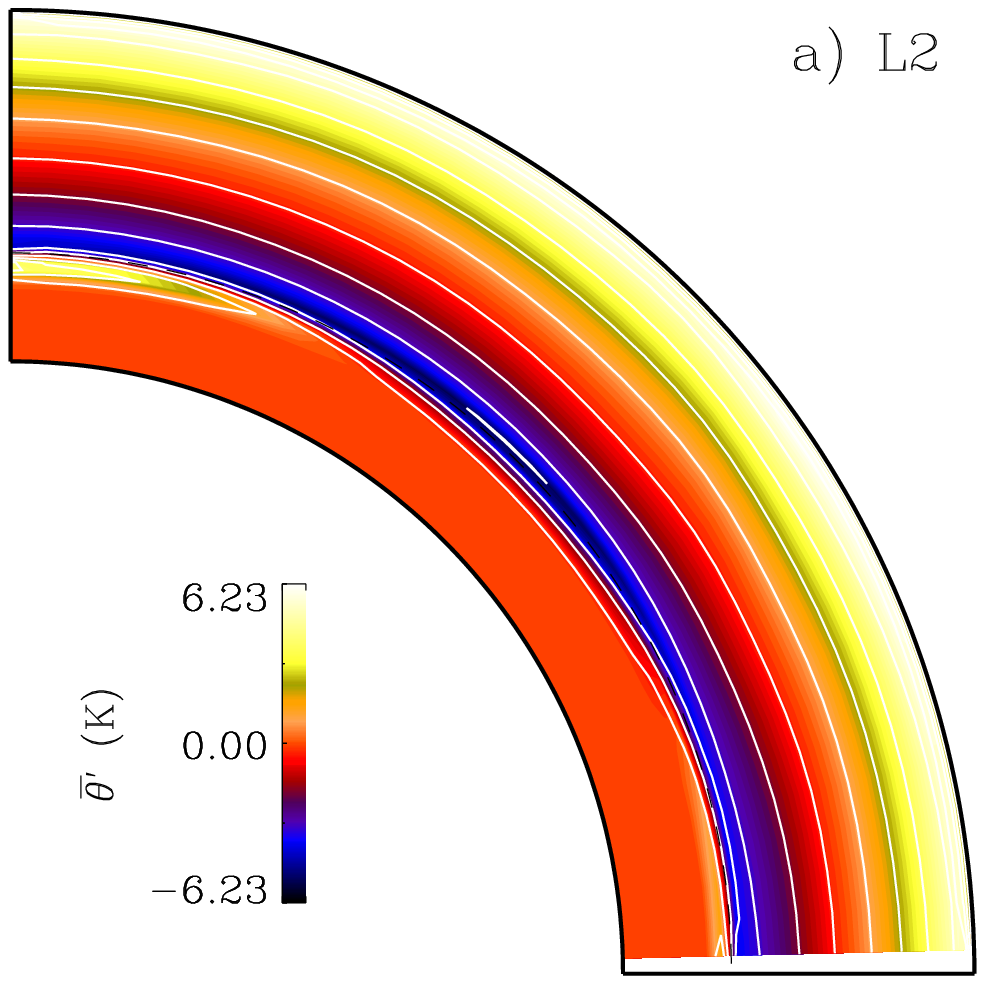}
\includegraphics[width=0.32\columnwidth]{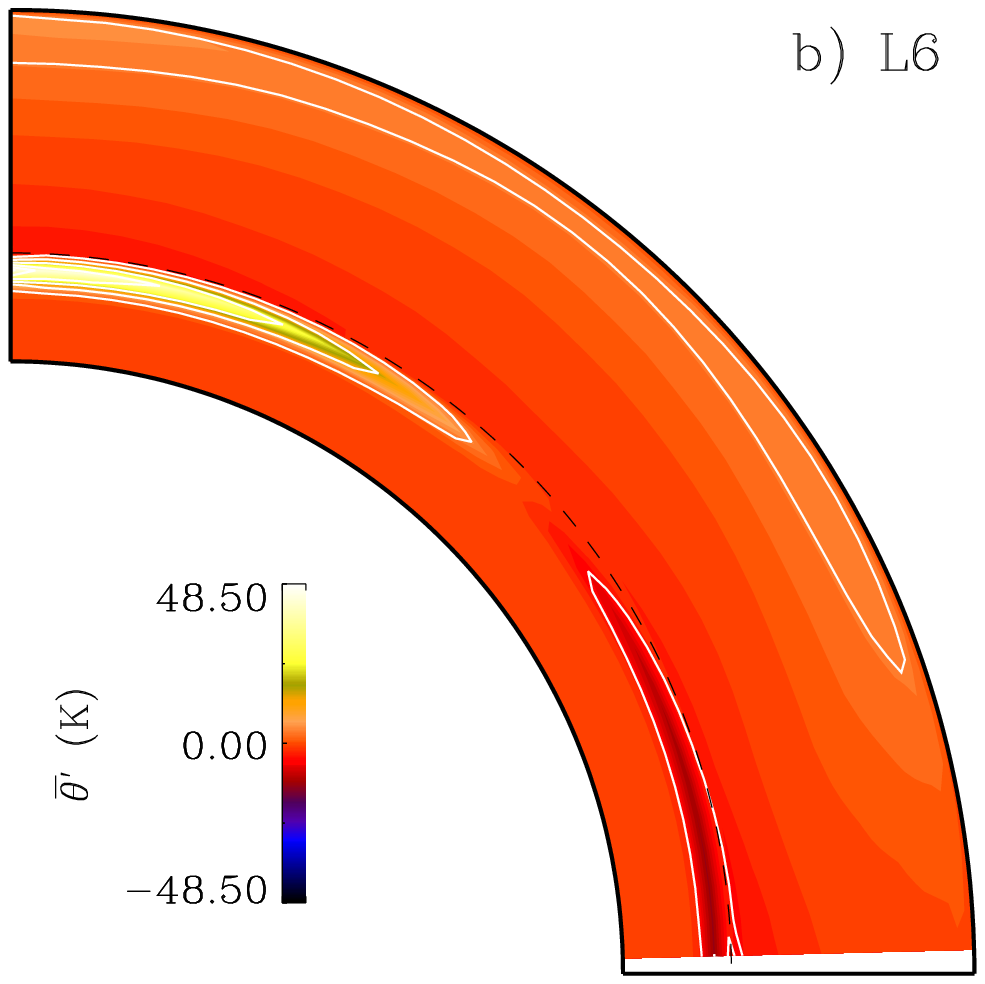}
\includegraphics[width=0.32\columnwidth]{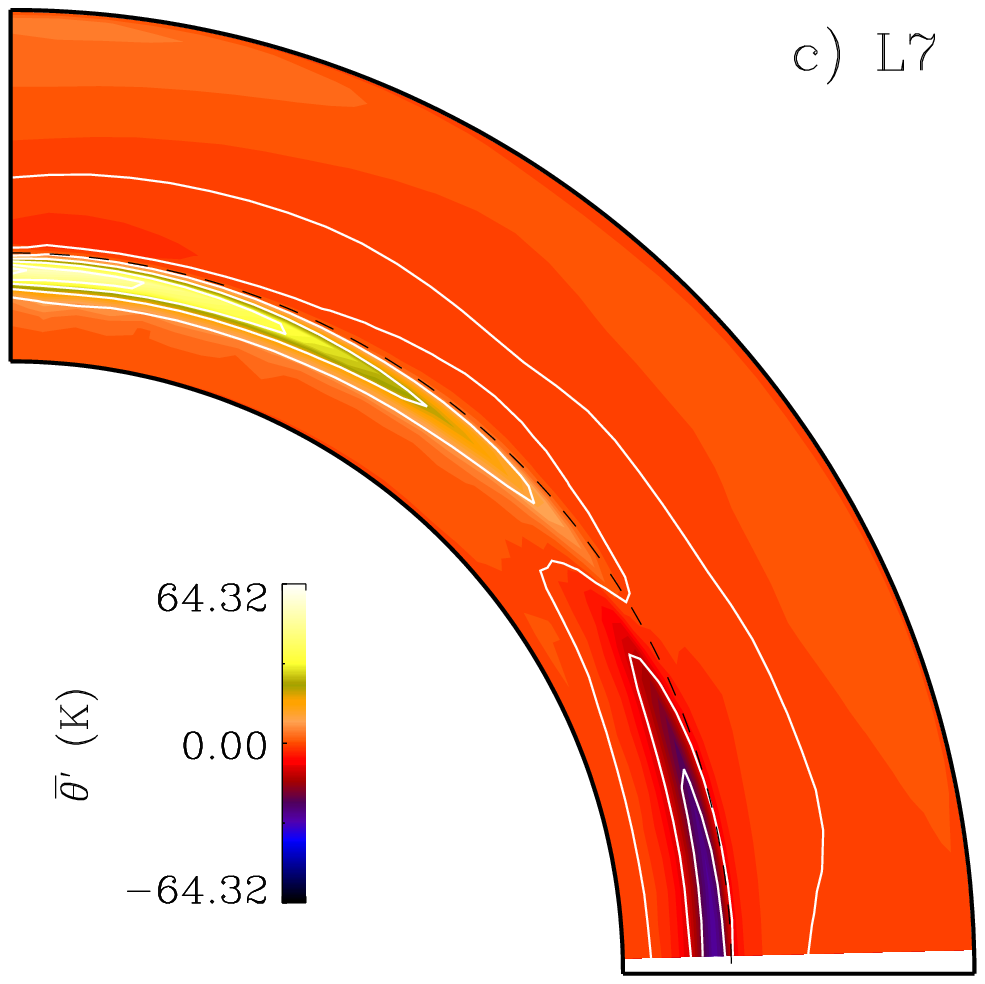}
\caption{Azimuthal average of $\Theta'$ for three different
rotation rates corresponding to models ~L2, ~L6 and ~L7.}
\label{fig.dthe}
\end{figure}
 
It is important to notice that
qualitatively the shape of the iso-rotation contours does not depend on
the dominance of the buoyancy or Coriolis forces; in all cases the
rotation contours are aligned along the rotation axis. To study
the role of baroclinicity in the resulting Taylor-Proudman
balance, in Fig. \ref{fig.dthe} we plot the azimuthal average of the
potential temperature fluctuation, $\mean{\Theta'}(r,\theta)$, for three
representative models: ~L2, ~L6 and ~L7.
Slow rotation models (L1-~L3) exhibit small variations of potential 
temperature
predominantly in the radial direction. In rapid rotating models (L4-~L7),
below the convection zone, $\mean{\Theta'}$ aquires larger values, 
it is positive at the 
poles and negative at the equator, indicating warmer poles.
However, within the convection zone, this latitudinal gradient does not
propagates and $\mean{\Theta'}$ varies mainly radially. The lack
of a latitudinal gradient in temperature explains why the
contours or differential rotation remain cylindrical (see Eq. \ref{eq.tp}).
The reasons and implications of this behavior will be explored
in a different paper. 

\begin{figure*}[h]
\includegraphics[width=0.32\columnwidth]{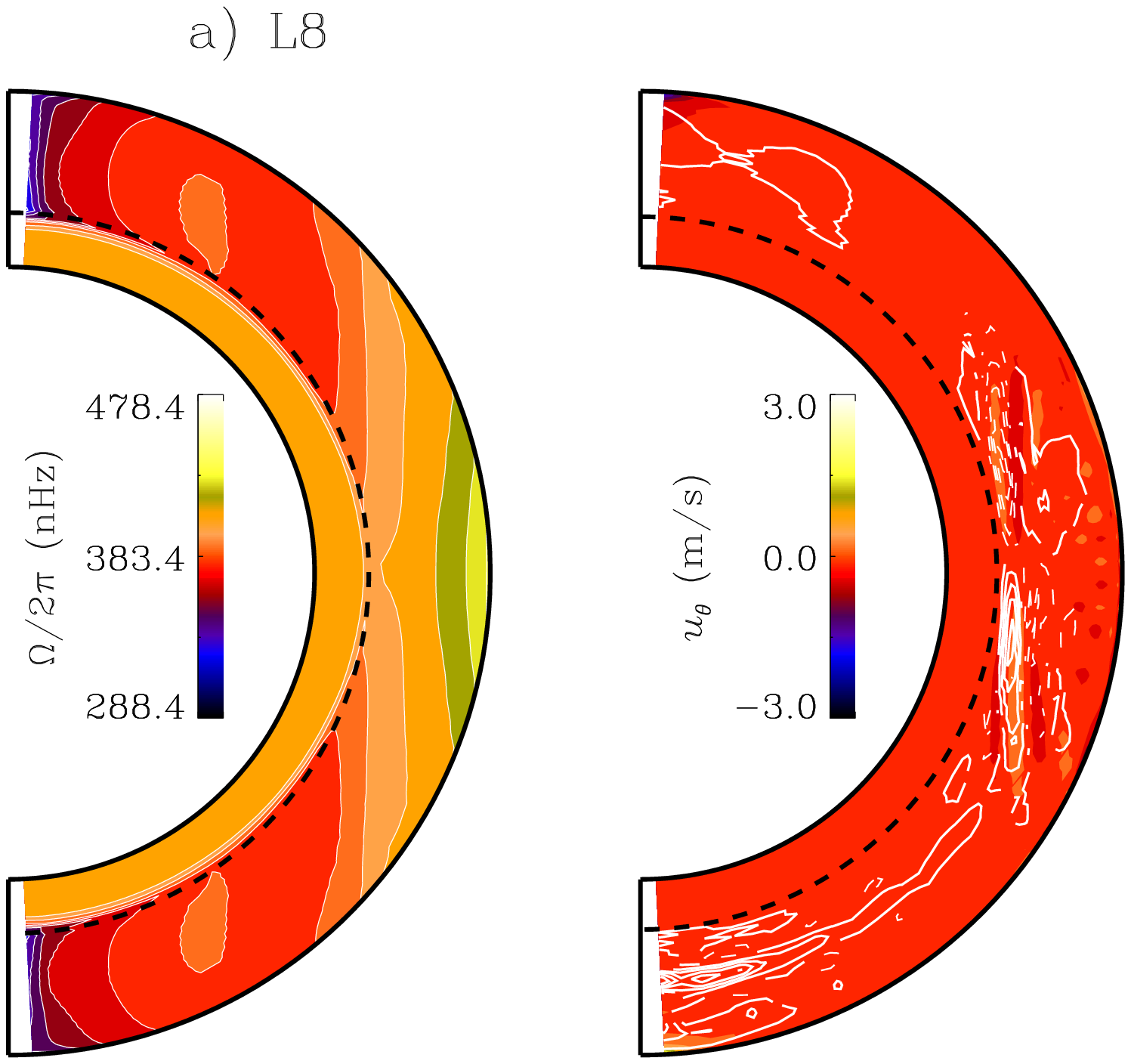}
\includegraphics[width=0.65\columnwidth]{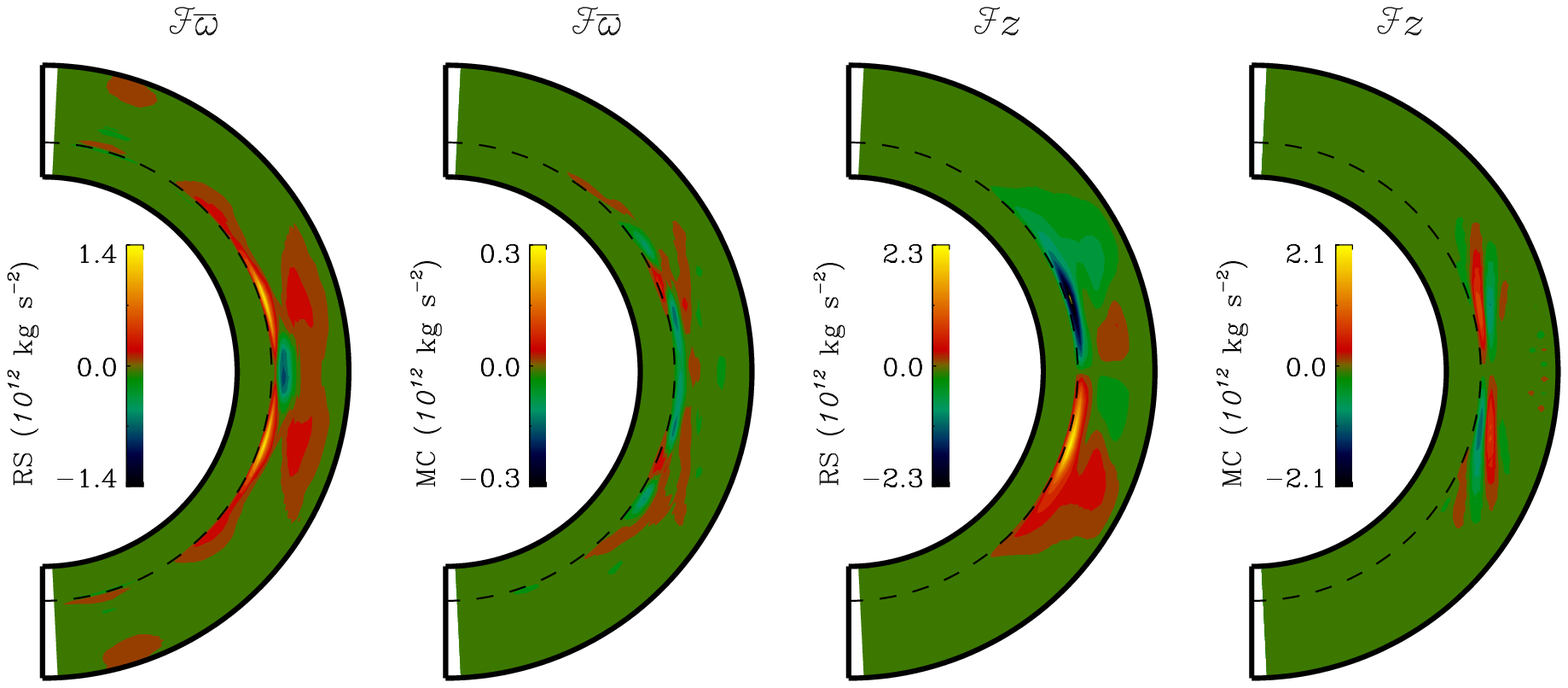}\\
\includegraphics[width=0.32\columnwidth]{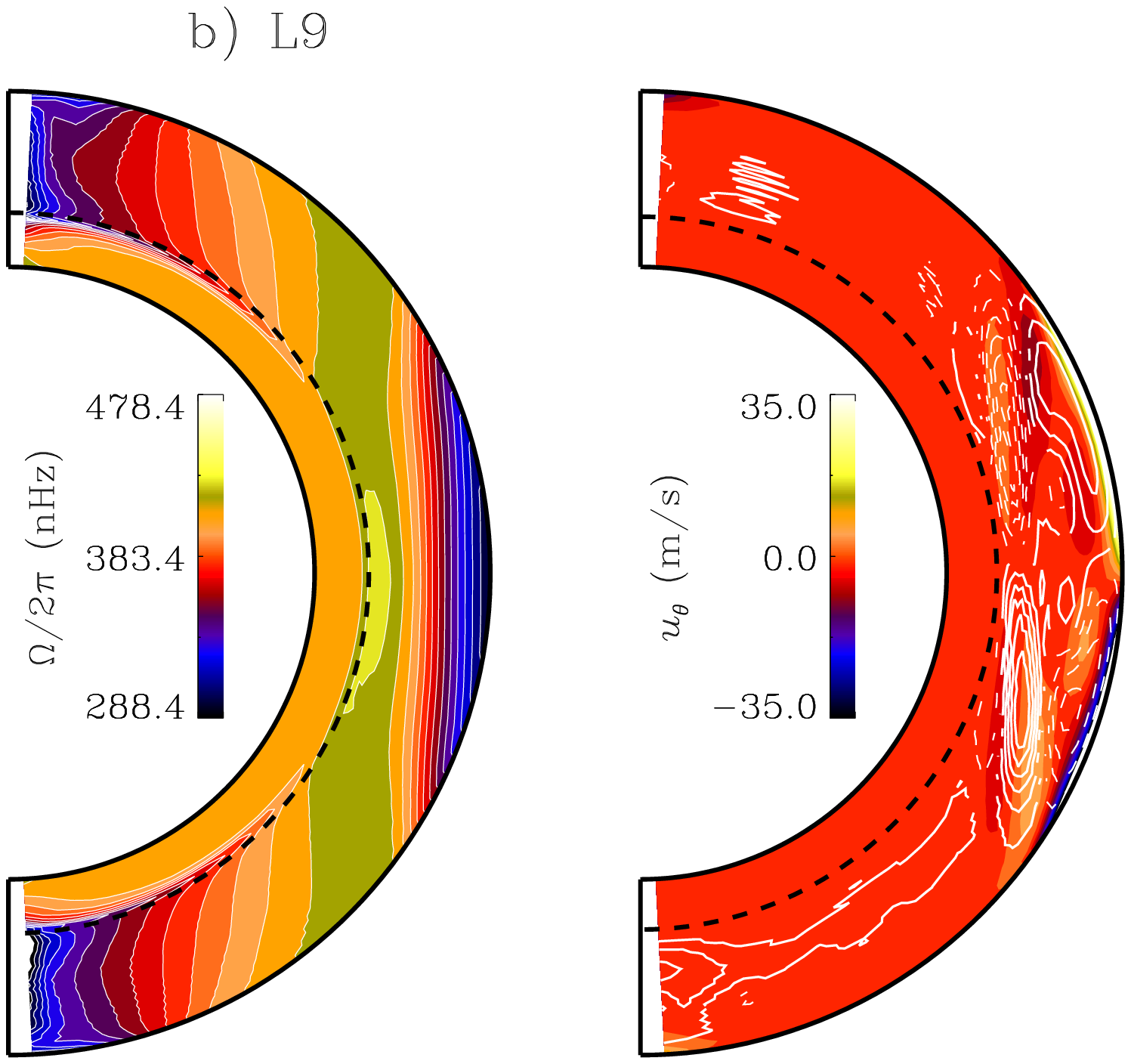}
\includegraphics[width=0.65\columnwidth]{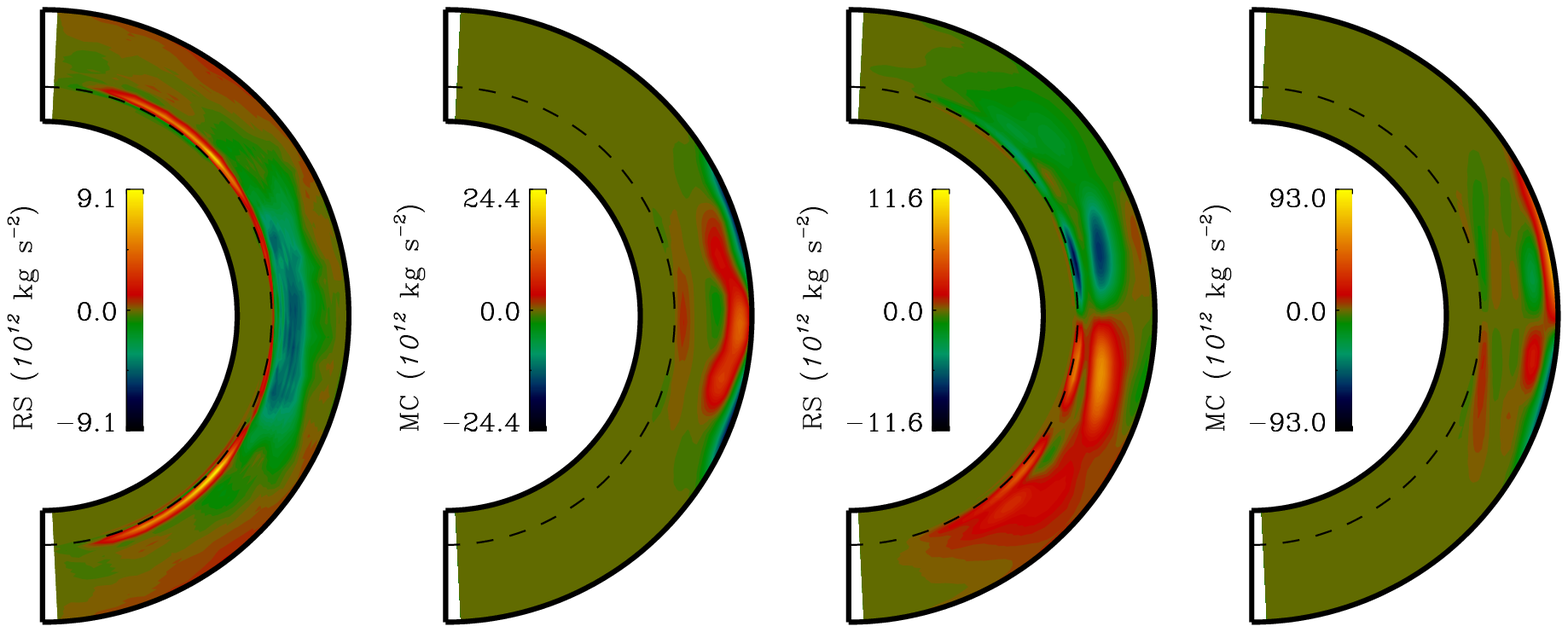}\\
\caption{Same than Fig. \ref{fig.df1} but for models a) ~L5 and b) ~L6. More 
(less) vigorous convection leads also to models with slower (faster) equator. The 
results are comparable to those of models ~L6 or ~L7 and ~L2,  
respectively. 
} 
\label{fig.df2}
\end{figure*}

\begin{figure}[H]
\includegraphics[width=0.98\columnwidth]{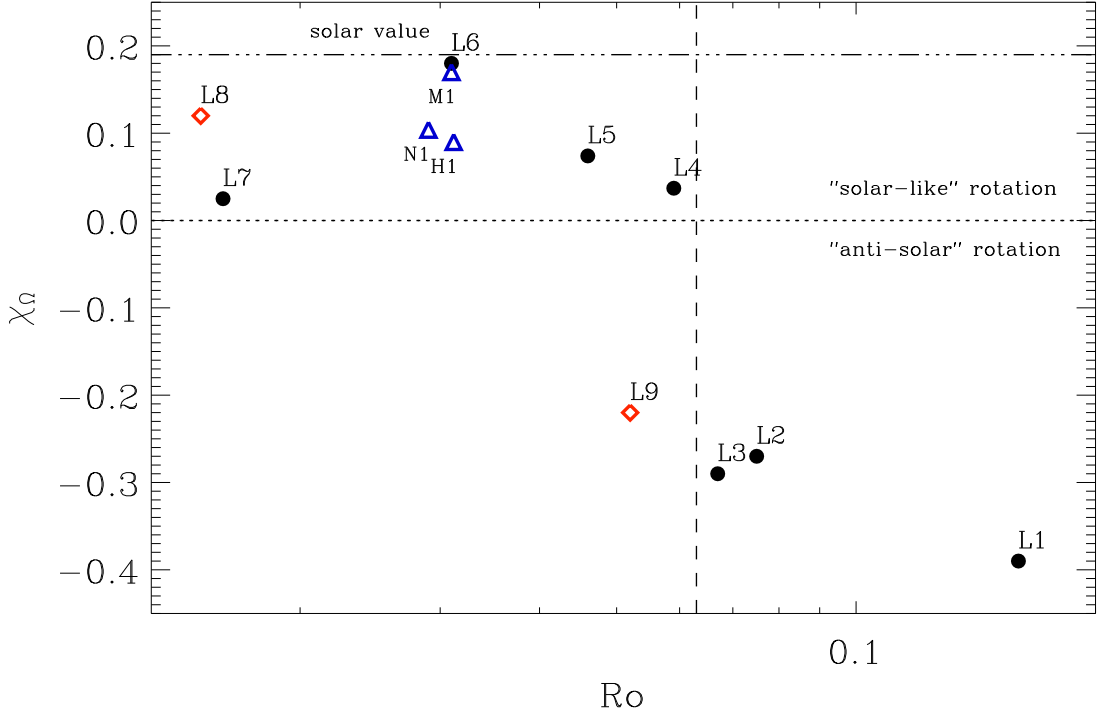}
\caption{Differential rotation parameter, $\chi_{\Omega}$ (Eq \ref{eq.chi}), as
function of the Rossby number, $\Ro$ for the simulation models defined in 
Table \ref{tbl.1}. The dashed vertical line at $\Ro=0.063$ indicates the sharp
transition between "solar"-type (prograde) and "antisolar" (retrograde) regimes
of the differential rotation. The dashed-dotted line indicates the solar value
of  $\chi_{\Omega}$.}
\label{fig.rovsro}
\end{figure}

The transition between the solar and anti-solar rotation regimes
(faster or slower equator) has been obtained since the early models
of \cite{Gi76} where he compared the Rayleigh with the Taylor 
numbers. In his Fig. 6 he depicts the regimes for transition 
between solid rotation, equatorial acceleration and high latitude
acceleration. Similar results for the differential rotation and
meridional circulation in the context of rotation of gas planets 
were recently reported by \cite{GWA13}. Performing simulations 
for different rotation
rates they found a sharp transition from the prograde
to retrograde rotation of equatorial regions. 
They compared the Rossby number at the equator with the Rayleigh
number at a mid depth, $\Ra^*$, and found that this transition
occurs at  $\Ra^*\sim1$. 
In this paper, we compare the differential rotation parameter,
defined  in Eq. (\ref{eq.chi}) with the Rossby number as it is
depicted in  Fig. \ref{fig.rovsro}. 
For the models with the same ambient state we 
find that the sharp transition between the prograde and retrograde
regimes occurs at $\Ro\simeq0.063$ ($\Ra^* \simeq 3$ in Table \ref{tbl.1}), 
indicated by the vertical dashed line in  Fig. \ref{fig.rovsro}. 
In models with the same rotation rate but
with higher (lower) superadiabaticity (models ~L8 and ~L9), the 
transition can happen at lower (higher) values of $\Ro$.  
It is noteworthy to mention
that the model ~L6 is in a good agreement with the solar value of
$\chi_{\Omega}$, indicated by the horizontal dashed-dotted line. 
For the model with faster rotation, ~L7, the differential rotation 
parameter, $\chi_{\Omega}$, decreases. Here it happens due to 
the formation of a cylinder of slow rotation at intermediate latitudes. 
However, similar global simulations of fast rotating stars also show 
a decrease in the latitudinal differential rotation \citep{BBBMT08,DCh13}.
\cite{KMGBC11} also explored the transition from anti-solar to solar-like
differential rotation in compressible simulation of convection in spherical 
wedges. In their Fig. 17 they compare the rotation parameter with the Coriolis 
number (inverse to $\Ro$). Unlike our results and those in \cite{GWA13},
in their case this transition seems to be smoother. 
However, due to all these differences in the model and experimental 
design this comparison remains inconclusive and warrants further studies.

\subsection{Convergence test}
\begin{figure}[H]
\includegraphics[width=0.6\columnwidth]{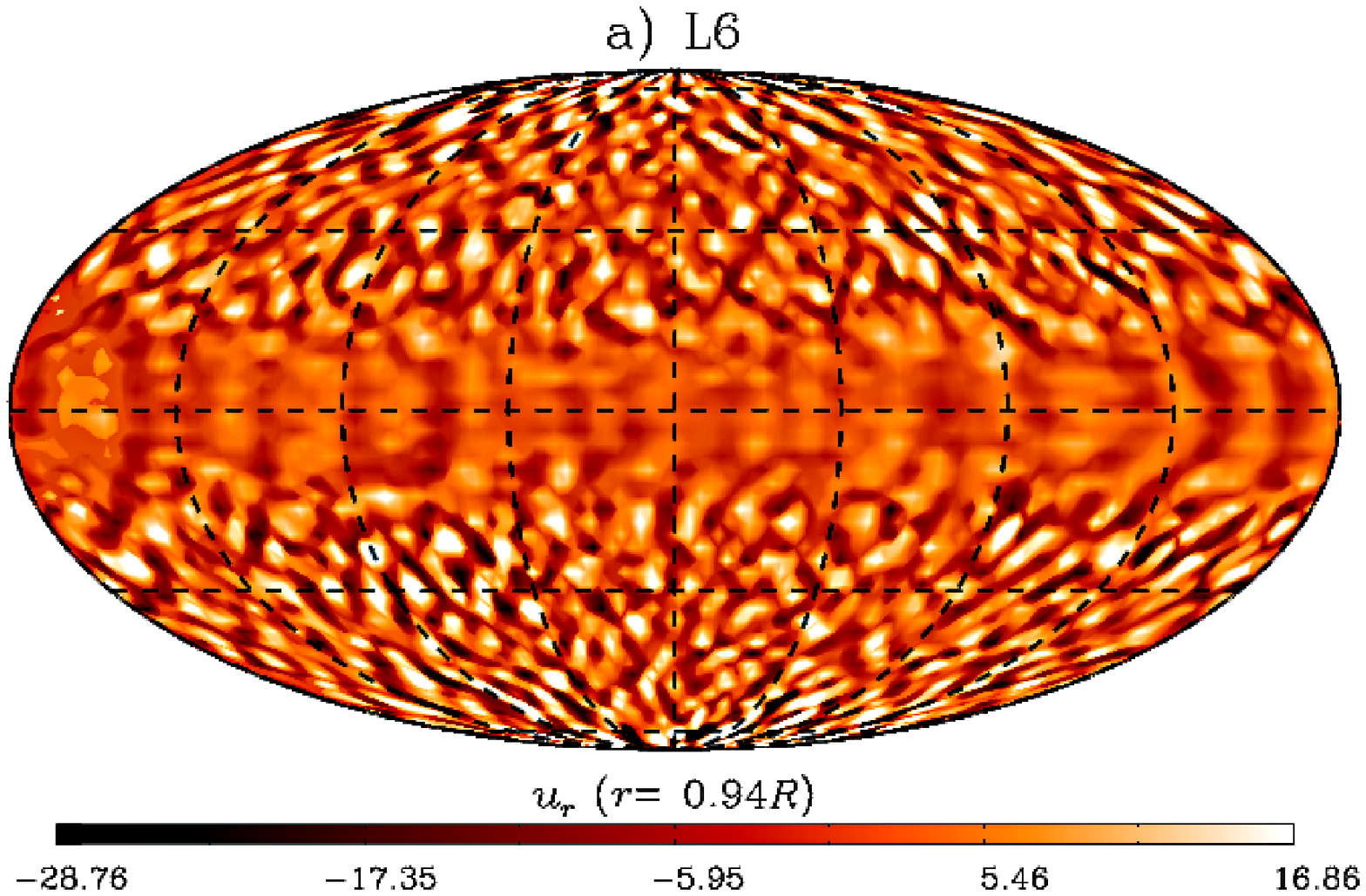}\\
\includegraphics[width=0.6\columnwidth]{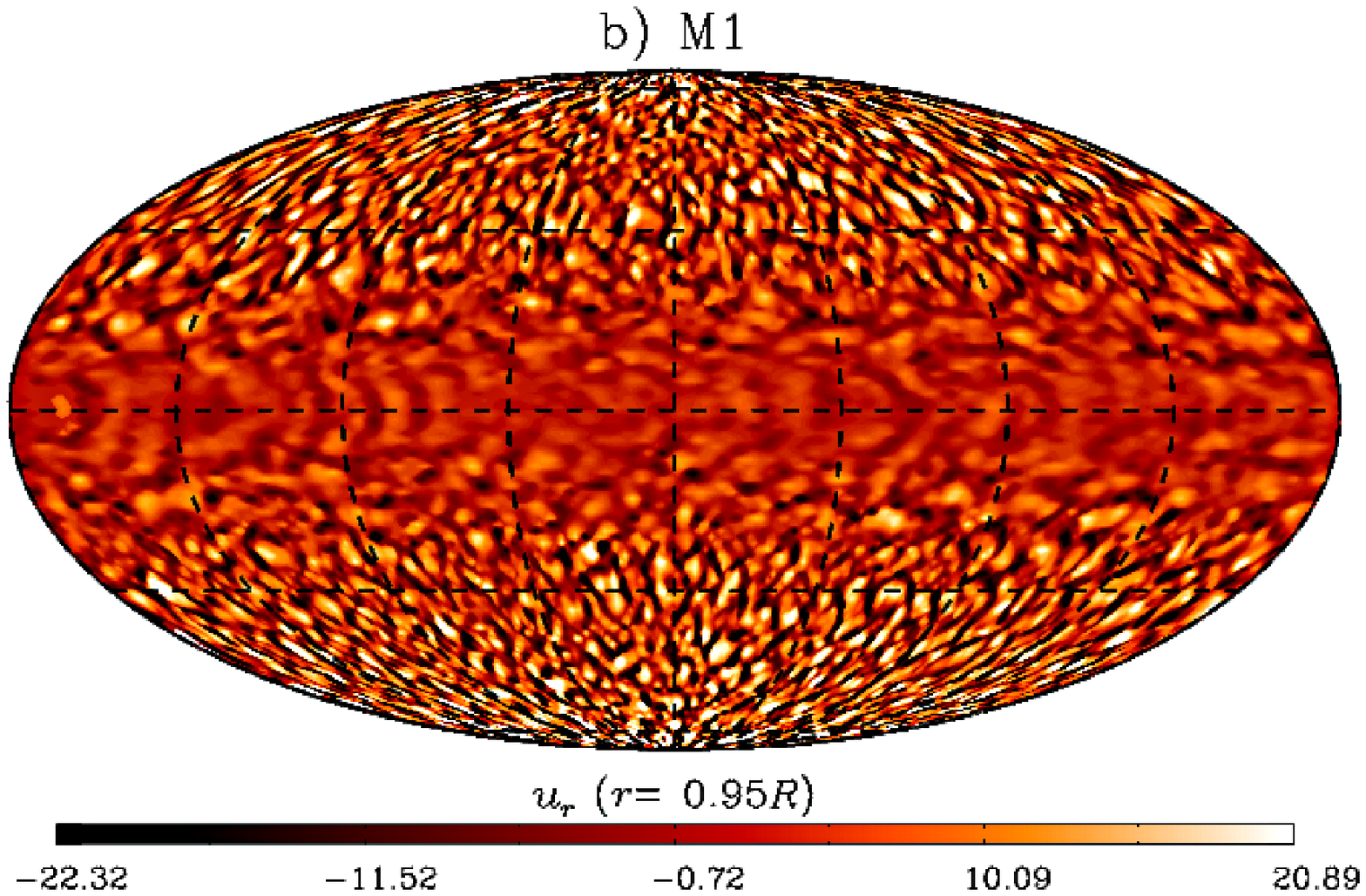}\\
\includegraphics[width=0.6\columnwidth]{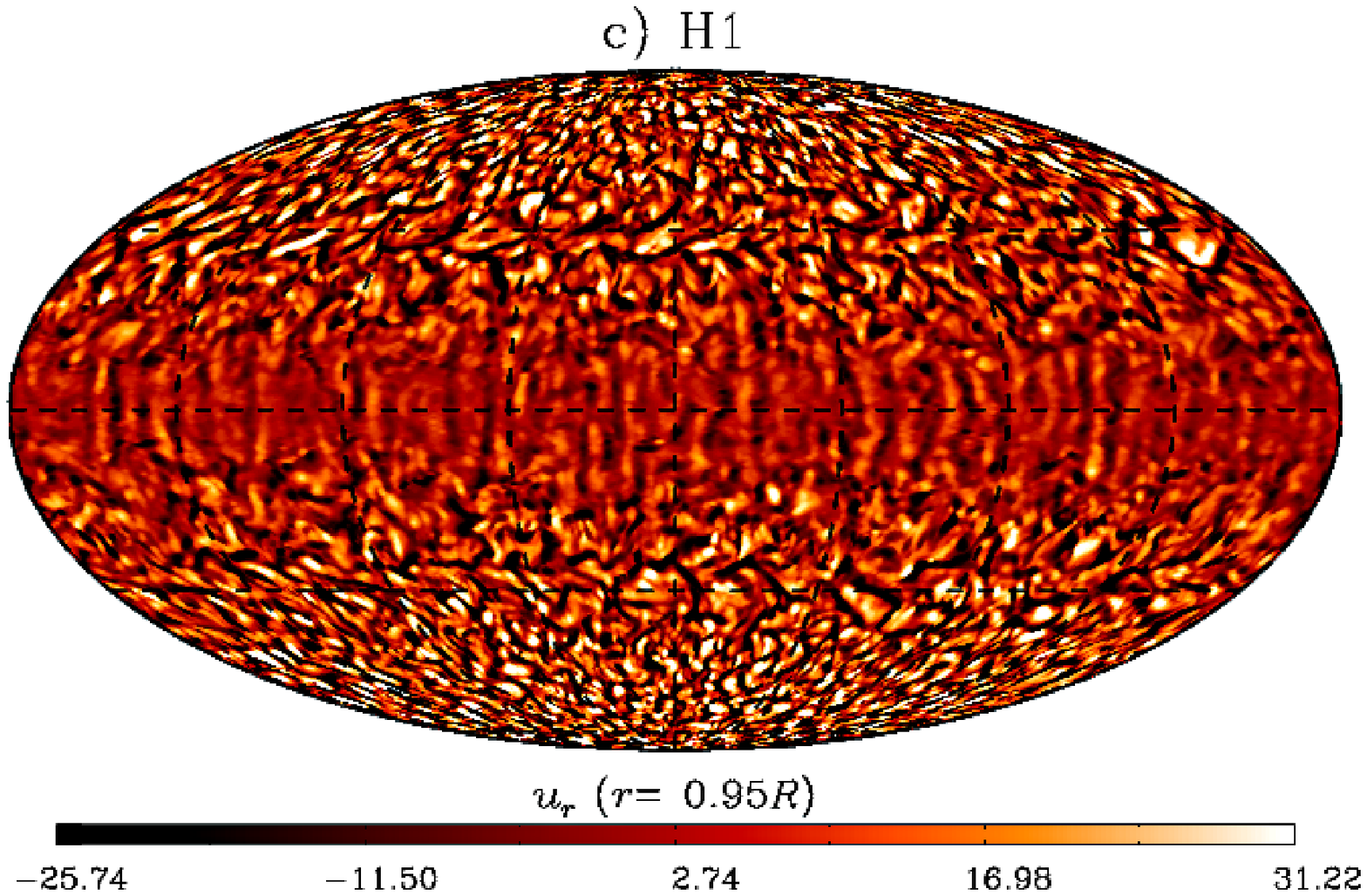}\\
\caption{Snapshots of vertical velocity, $u_r$ for models ~L6, ~M1 
and ~H1 (Table \ref{tbl.1})  with
lower, intermediate and higher grid resolution (from top to bottom).}
\label{fig.ur}
\end{figure}

As a test for the implicit SGS model, we performed a set of simulations 
presented for the same stratification ($\Theta_e$) and the same
rotation rate ($\Omega_{\odot}$) but for different mesh resolutions.
Run ~L6, presented in the previous section was for a relatively coarse grid
model with $128\times64\times47$ mesh points. Models M1 and
L1 have $2$ and $4$ times higher resolution, respectively. Snapshots 
of the vertical velocity for the cases ~L6, ~M1 and ~H1 are shown, from
top to bottom panels, in Fig. \ref{fig.ur}.  
In all three cases convective ``banana" cells elongated along the rotation
axis appear  in a belt of 
$\pm30^{\circ}$. However, the scale of these banana cells depends
on the resolution, i.e., a larger number of azimuthal modes seems to 
be excited for higher resolutions. The higher latitudes are populated with 
smaller and more symmetrical convection cells whose horizontal scale 
also decreases with the increasing of the resolution. 

\begin{figure}[H]
\includegraphics[width=0.98\columnwidth]{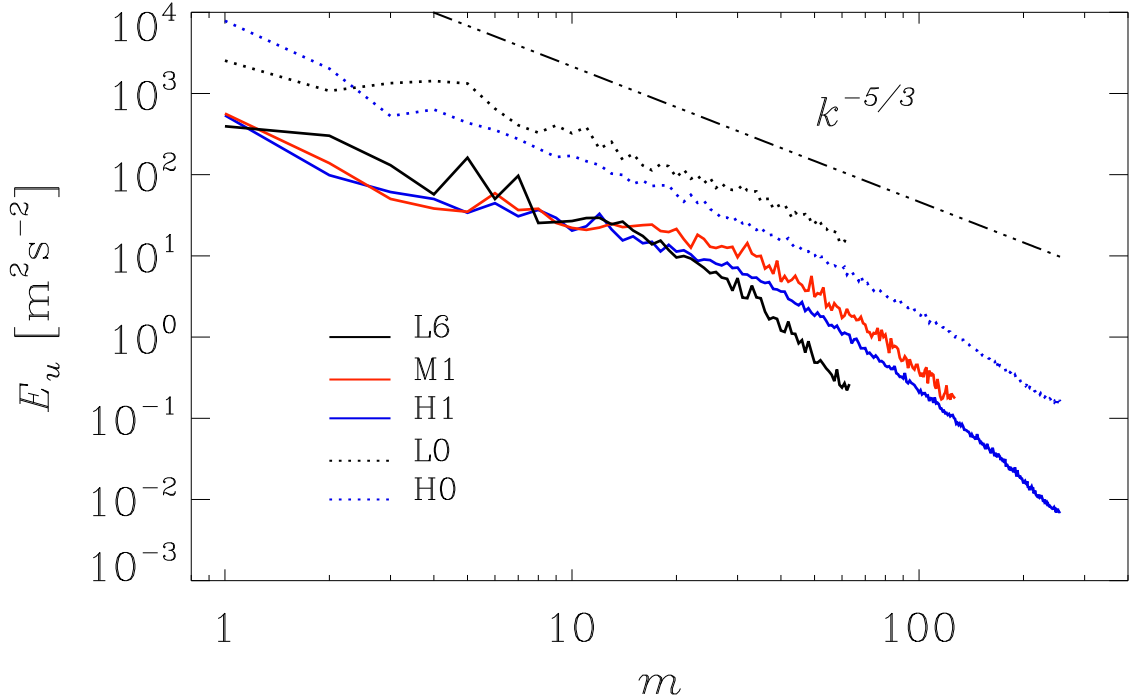}
\caption{Power spectrum of the velocity field at $r=0.95\Rs$ as a 
function of the azimuthal wavenumber, $m$,  for simulations with 
different numerical resolution, ~L6, ~M1 and ~H1
(see legend and Table \ref{tbl.1}).
Continuous (dotted) lines correspond to the models with 
(without, models ~L0 and ~H0) rotation. 
The dot-dashed line indicates the $-5/3$ 
power decay.}  
\label{fig.pe}
\end{figure}

To study the energetics of the interacting large and small 
scales for each of these simulations
we have computed the power spectrum of the kinetic energy at 
a given radius, using the Fourier transform in 
the azimuthal direction:
\begin{equation}
{\hat u}_m=\int {\bm u}(r_t,\theta,\phi) \exp (im\phi)\frac{d\phi}{2\pi},
\end{equation}
where $m$ is the azimuthal wave number, and the power spectrum 
definition \citep{MTBM09}:
\begin{equation}
E_u=\brac{{\hat u}_m^2}_{\theta},
\end{equation}
where the brackets denote average over latitude. 
Note that $\sum{E_u}=\brac{{\bm u}^2}$. The spectra for
models ~L6, ~M1 and ~H1 are plotted in Fig. \ref{fig.pe}. For the 
coarser resolution (continuous black line) the spectrum shows an 
energy decay from the smaller
wavenumber up to $7\lesssim m \lesssim 15$. There are peaks of power 
at $m\simeq5$ and $m\simeq7$, which 
probably correspond to the scales of the banana cells. From $m\simeq15$
to the smallest resolved scale, $m=64$, there are two different
sub-ranges, one of which is compatible with the inertial turbulent decay 
(although not exactly following the Kolmogorov $k^{-5/3}$ law, see 
dot-dashed line) up to $m\simeq30$, 
followed by a faster decay.  In total, the energy spectra spans $\sim3$ 
orders of magnitude. The spectrum for the run M1 (red line) is 
similar to ~L6,
but in this case the excess of power is at $m\simeq6$. There is a plateau
of energy from  $10\lesssim m \lesssim 20$ followed by a decay up to 
the smallest wave number $m=128$.  The model H1 with higher resolution 
(blue line) does not have a plateau, and the energy seems to decay 
continuously, with
a seemingly inertial sub-range between $12\lesssim m \lesssim 30$.
The peak at $m\simeq12$ contains the energy of thin banana cells
observed in this simulation. In total this model spans $\sim5$ orders
of magnitude in energy. For comparison we have plotted also the 
power spectrum of the models without rotation for low, ~L0, and high,
~H0, resolutions (black and blue dotted 
lines). Evidently the convective energy is higher in the models without
rotation.  It is interesting how this simulations show a
turbulent decay consistent with  $k^{-5/3}$ 
which spans several orders of magnitude in energy and for
a long range of scales. 

\begin{figure*}[h]
\includegraphics[width=0.36\columnwidth]{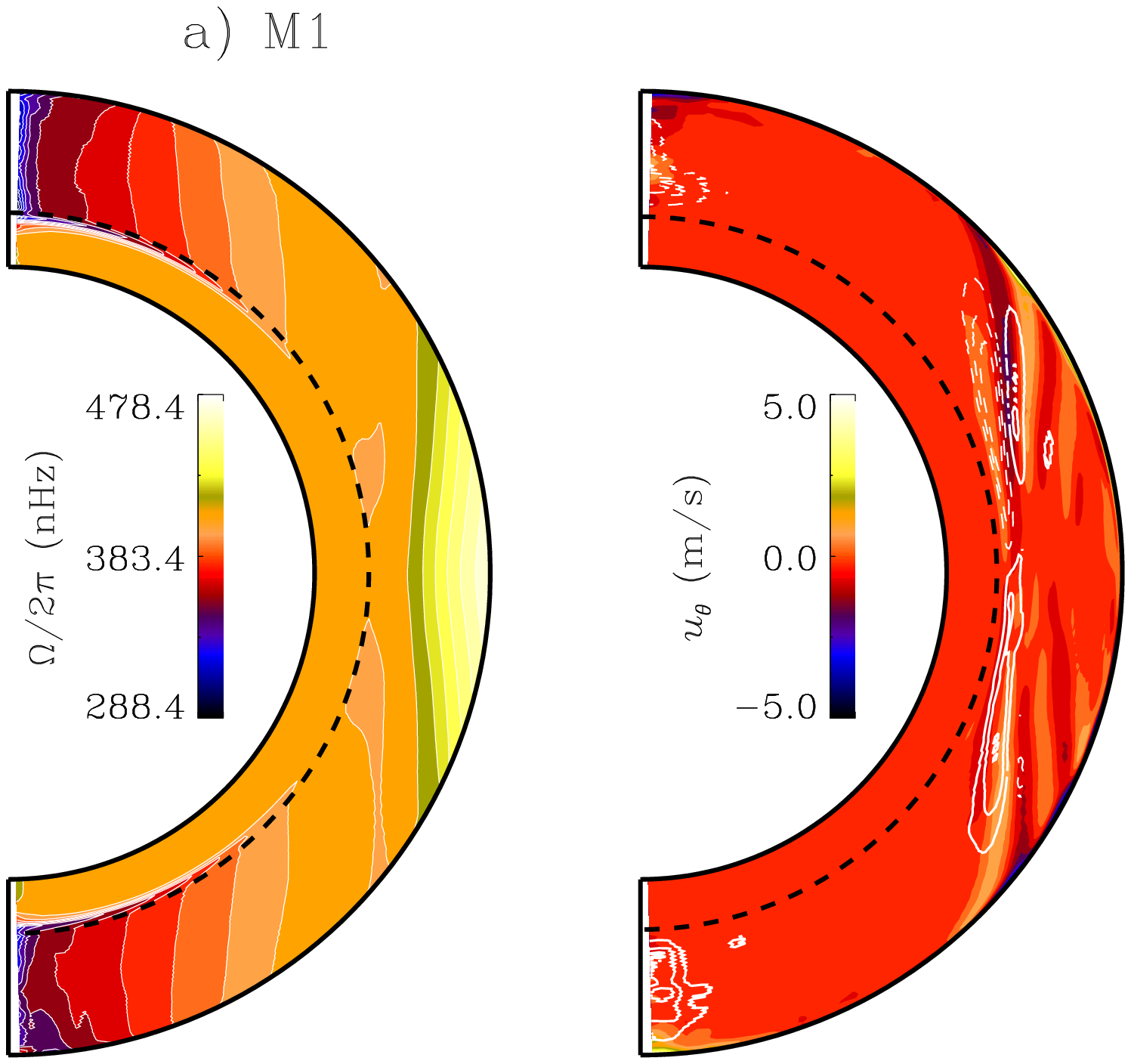}
\includegraphics[width=0.7\columnwidth]{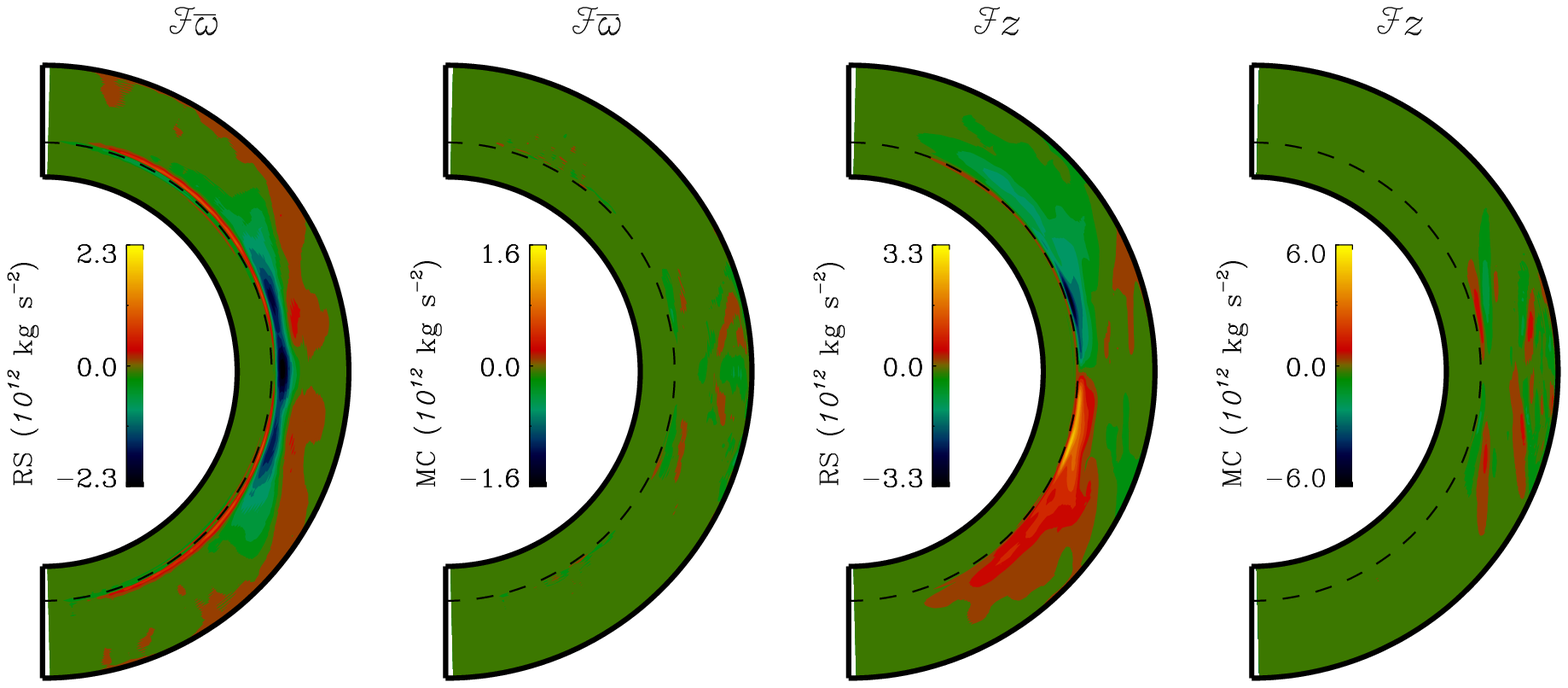}\\
\includegraphics[width=0.36\columnwidth]{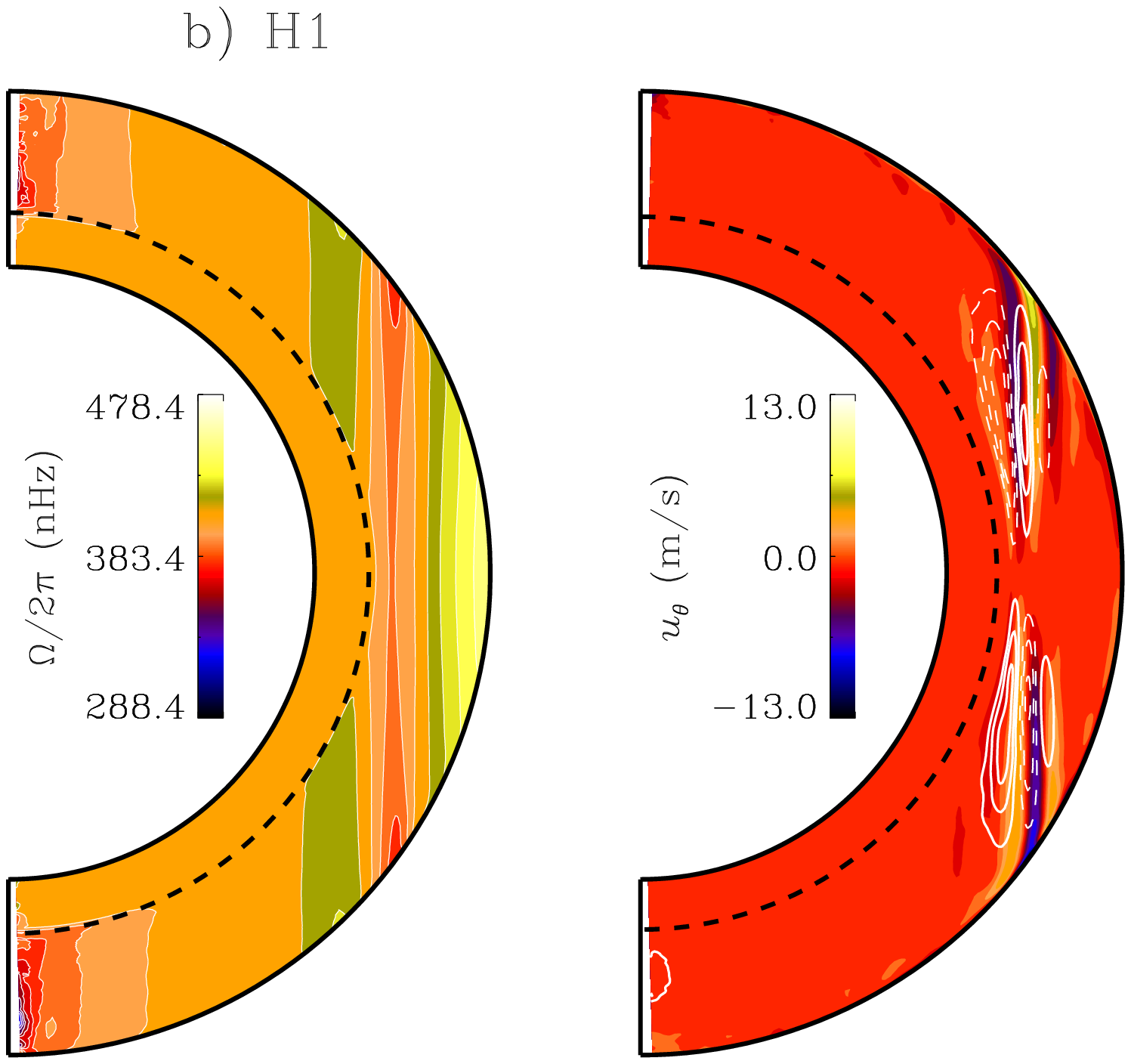}
\includegraphics[width=0.7\columnwidth]{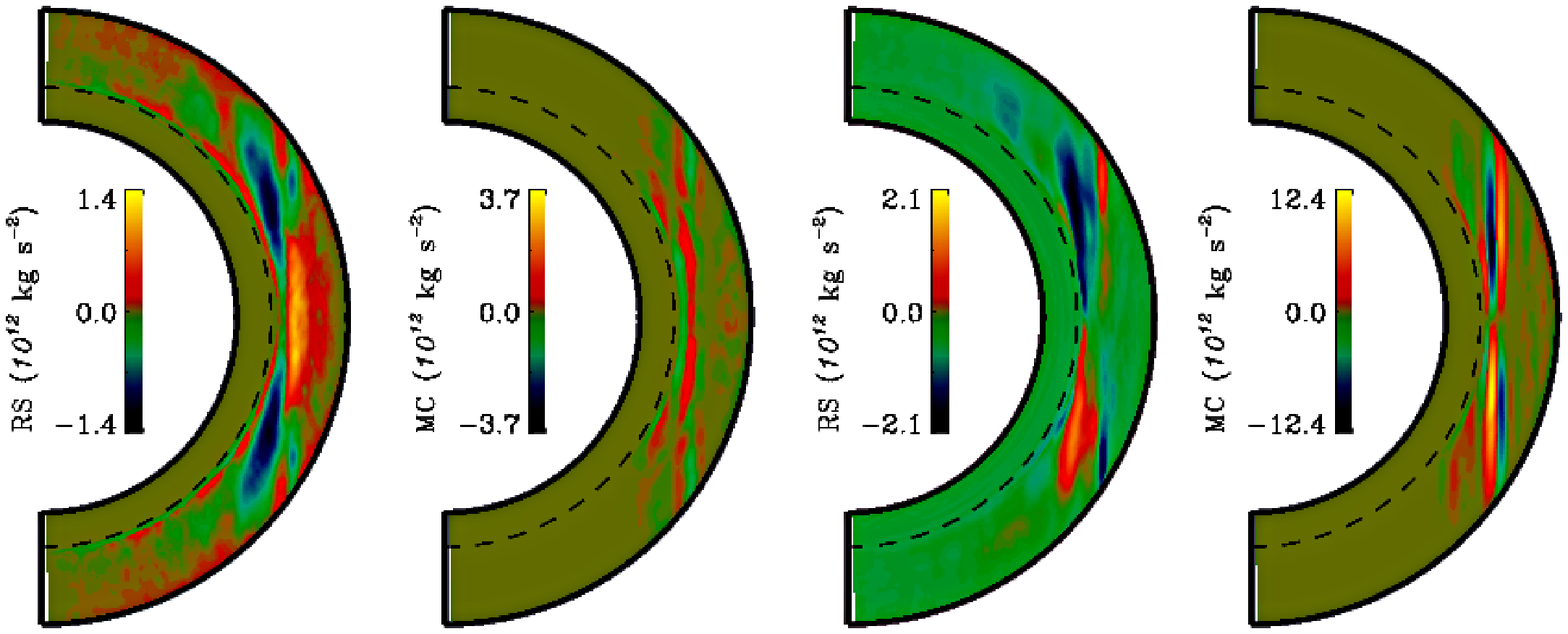}\\
\caption{Same than \ref{fig.df1} but for models ~M1 and ~H1 with 
$256\times128\times94$ and $512\times256\times188$ mesh points,
respectively.}
\label{fig.df3}
\end{figure*}

Similarly to Fig. \ref{fig.df1}, Fig. \ref{fig.df3} shows 
the profiles of differential rotation and meridional circulation
obtained for models ~M1 and ~L1, as well as the angular
momentum flux components.  In model ~M1 the mean rotation and 
meridional flows slightly resemble 
those of model ~L6. In this case, however, the rotation shows a 
band of rotation with the speed of the stable layer, which spans 
over a large latitudinal extent. Both quantities, the Rossby number, 
$\Ro$, and the rotation parameter, $\chi_{\Omega}$  are similar to those of 
model ~L6, so that this case appears 
almost at the same position in Fig. \ref{fig.rovsro}. Regarding the 
angular momentum fluxes,  the meridional 
distribution of ${\cal F}^{RS}_{\varpi}$  and ${\cal F}^{RS}_z$ is
qualitatively similar to ~L6. In the case ~M1, the Reynolds stress 
contribution is about one half of that in case ~L6 
(see Fig. \ref{fig.df3}a).  On the other hand,  
${\cal F}^{MC}_{\varpi}$  and ${\cal F}^{MC}_z$ increase
by a factor of two. Given the resemblance in the Reynolds stresses profiles,
the differential rotation  does not differ abruptly
between cases ~L6 and ~M1. 
The meridional motions are, however, more pronounced in this case.
Like in case ~L6, there are several cells in the meridional flow profile
elongated in the $z$ direction. 

This contribution of meridional circulation to the angular momentum flux
appears further enhanced in the high resolution model,
 ~H1 (Fig. \ref{fig.df3}b). Although there is an
equatorial acceleration due to a positive ${\cal F}^{RS}_{\varpi}$ flux, 
both ${\cal F}^{RS}_{\varpi}$ and
${\cal F}^{RS}_{z}$ are decreased in amplitude. Most importantly, 
there is no morphological similarity to the cases ~L6 and ~M1. This leads
to a different profile of  $\Omega$. Instead of a continuous decrease
with latitude, there is a retrograde jet
which appears at the surface between $~20^{\circ}$ and $~40^{\circ}$.
This seems to be the consequence of meridional motions aligned  
along the tangent cylinder at $r\simeq0.71\Rs$. Another prograde jet 
appears at mid latitudes, from there up to the poles
the azimuthal velocity equals the rotation rate of the frame,
with a slight decrease towards the poles. 
In this case the meridional motions enhance
${\cal F}^{MC}_{\varpi}$ and ${\cal F}^{MC}_{z}$ by another 
factor of 2.
 
It is surprising that although the Rossby number, $\Ro$, is similar
for the models with three different resolutions, indicating that the 
balance between the Coriolis and buoyancy forces is roughly the same,  
in case ~H1 the value of $\chi_{\Omega}$ ($\simeq0.09$) is smaller than
in the lower resolution models ~L6 and ~M1, while
${\cal F}^{MC}$ is larger. It might be the case that when a more complex
and turbulent flow develops, small-scale motions contributing to the angular
momentum transport diffuse on shorter timescales so that the Coriolis force 
does not affect them. The appearance of these new scales of motion might
also change the role of the sub-grid scale transport, relegating its
action to scales not resolved.  The situation could be that the 
smallest resolved scales are already modifying the system. The question
that arises, and deserves further investigation, is why these effects
are not captured by the sub-grid scale transport in the coarser 
resolution cases? It is puzzling why our model ~L6 with the rather
coarse resolution reproduces the solar rotation closer than the higher
resolution models, ~M1 and ~L1. Perhaps it may be necessary to include 
explicit eddy transport in the higher resolution models (i.e.,
include in the equation an explicit term for the turbulent heat
diffusion or for the viscosity) in order to increase the efficiency
of the unresolved turbulent transport. In other words, this will modify
the effective Prandtl number of the simulation. Another possibility 
could be to readjust the parametrized turbulent heat flux 
profile, $\Theta_e$.

\subsection{The solar near-surface shear layer}
\label{sec.nssl}

In the upper part of the solar convection zone the 
velocities are such that the turnover time is of the order of 
minutes (for granulation) or hours (super-granulation). These 
time scales
are smaller when compared with the solar rotation period, $28$ days,
indicating that there the buoyancy force is dominating over the Coriolis
counterpart. Observations indicate that in this region the radial shear 
is negative, i.e., the 
angular velocity decreases with radius. In the simulations described in
the previous sections, an 
inward angular momentum flux is clearly observed 
in the buoyant dominated regime.  It corresponds to an angular 
velocity which decreases radially outwards. In order to simulate the
buoyancy dominated regime, we can take advantage of 
our formulation 
of the energy equation and impose a rapid decrease of the 
potential temperature in the upper part of the convection zone.   
Although radiative cooling is not considered in the simulations, this
sharp decline of the potential temperature (or entropy) occurs 
indeed at the upper part of the 
convection zone due to hydrogen ionization. 

\begin{figure}[H]
\includegraphics[width=0.98\columnwidth]{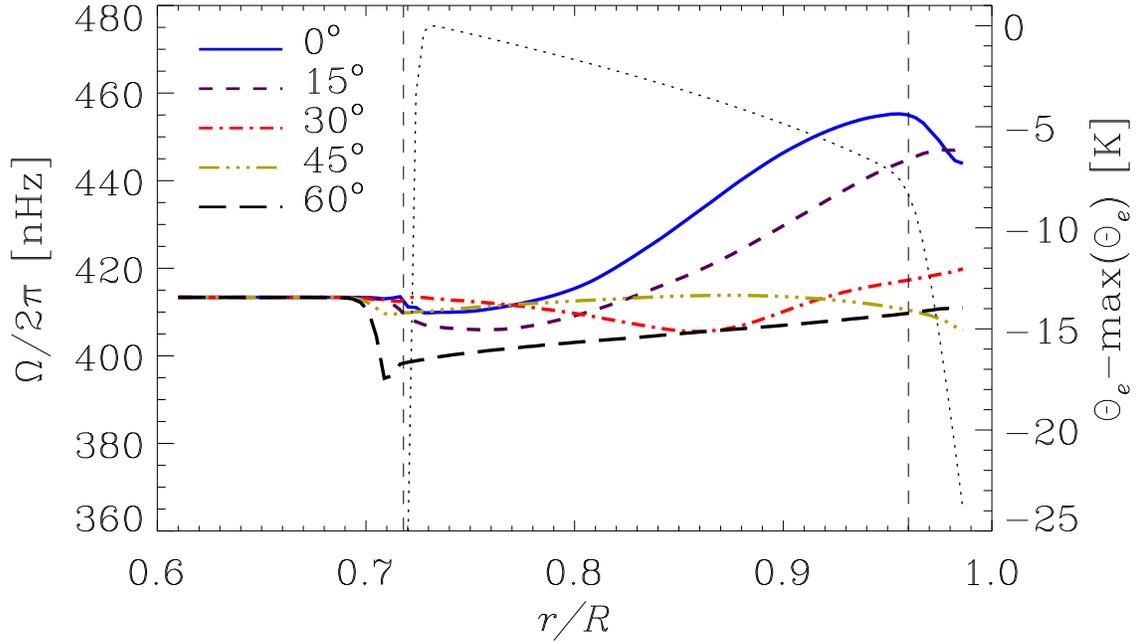}\\
\caption{Radial profile of the simulated angular velocity for model ~N1.
Different line styles and colors correspond to different latitudes as 
indicated in the legend. Dotted line indicates the radial profile
of the $\Theta_e$ (Eq. \ref{eq.pind2}). At equatorial latitudes the angular 
velocity decays at the upper part of the domain (blue line).}  
\label{fig.rpnssl}
\end{figure}

In the model labeled as ~N1 we extend the domain in the radial 
direction up to $r=0.985\Rs$ and modify the ambient state 
(Eq. \ref{eq.pind}) by considering the polytropic index 
\begin{eqnarray}
\label{eq.pind2}
m(r)=m_r &+& \Delta m \frac{1}{2}\left[1+ {\rm erf} \left( \frac{r-r_{tac}}
     {w_t} \right) \right] \\\nonumber
    &+& \Delta m_2 \frac{1}{2}\left[1+ {\rm erf} \left( \frac{r-r_{nssl}}
     {w_t} \right) \right] \;,
\end{eqnarray}
where $\Delta m_2=m_{nssl}-m_{cz}$, with
$m_{nssl}=1.4996$ and $r_{nssl}=0.96\Rs$ (see black dotted line in 
Fig. \ref{fig.rpnssl}).

\begin{figure}[H]
\includegraphics[width=0.98\columnwidth]{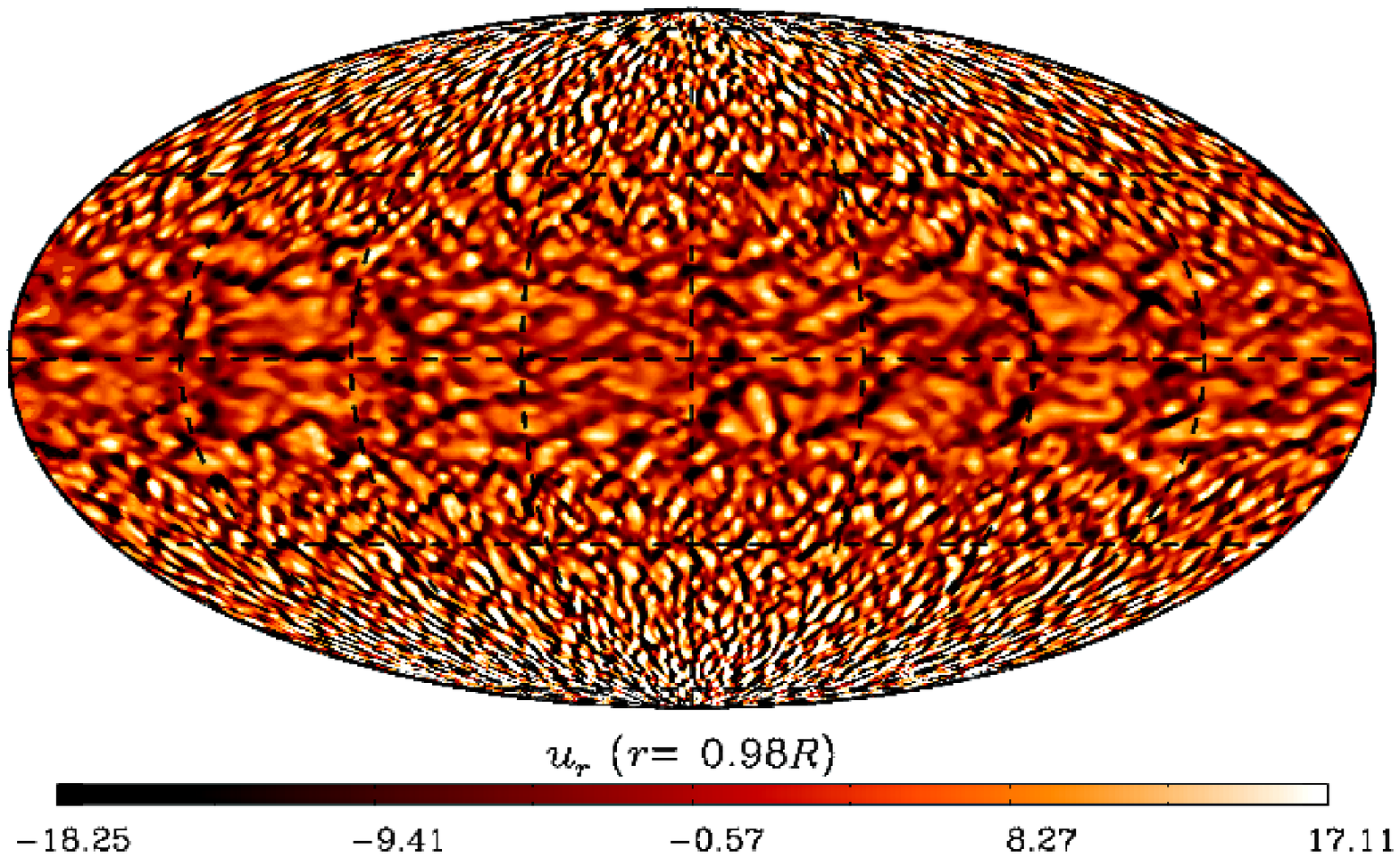}\\
\includegraphics[width=0.98\columnwidth]{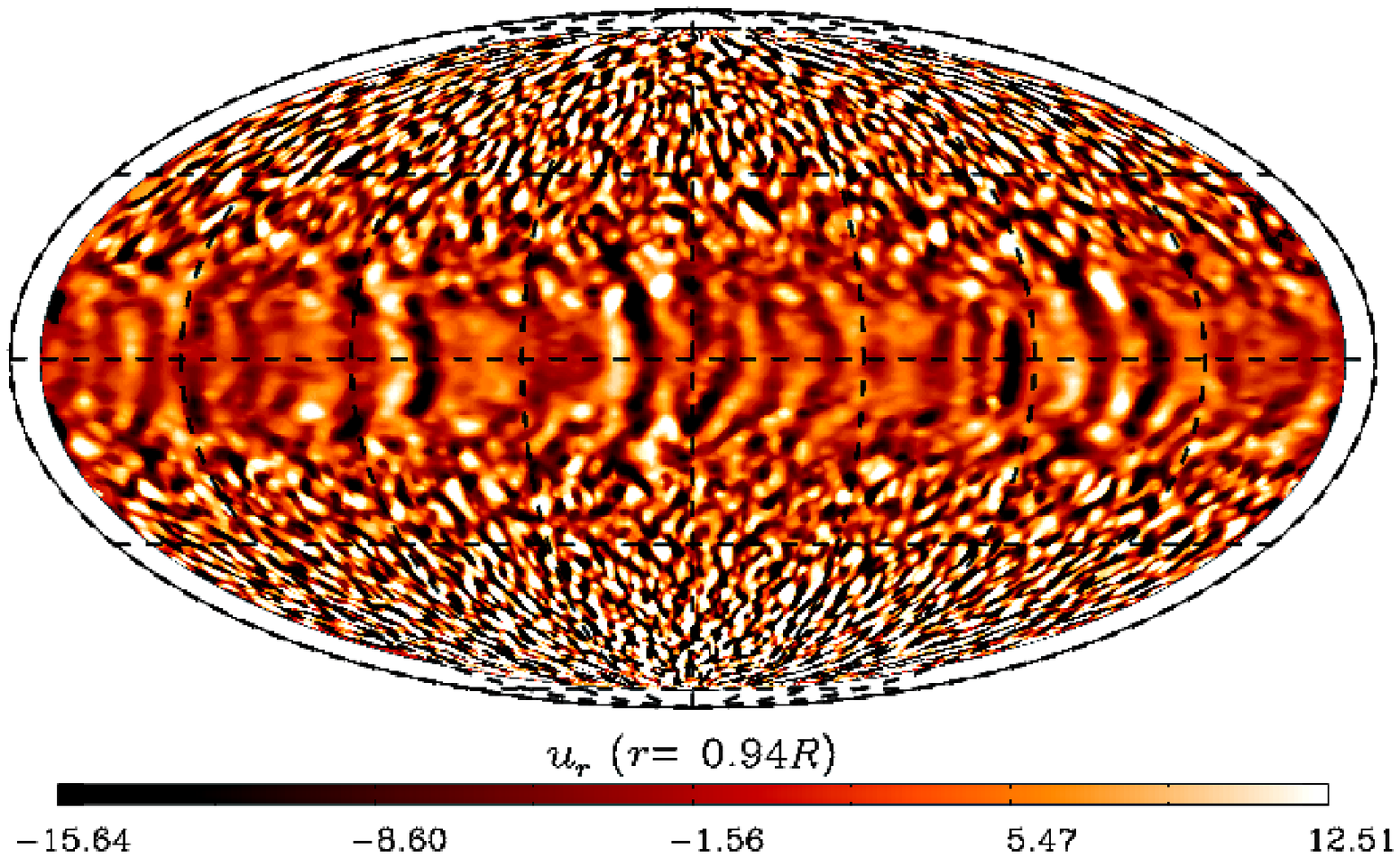}\\
\caption{Snapshots of vertical velocity, $u_r$ (in m/s), of the model ~N1. 
Upper and bottom panels correspond to $r=0.98\Rs$ and $r=0.94\Rs$.}
\label{fig.urns}
\end{figure}

\begin{figure}[H]
\includegraphics[width=0.98\columnwidth]{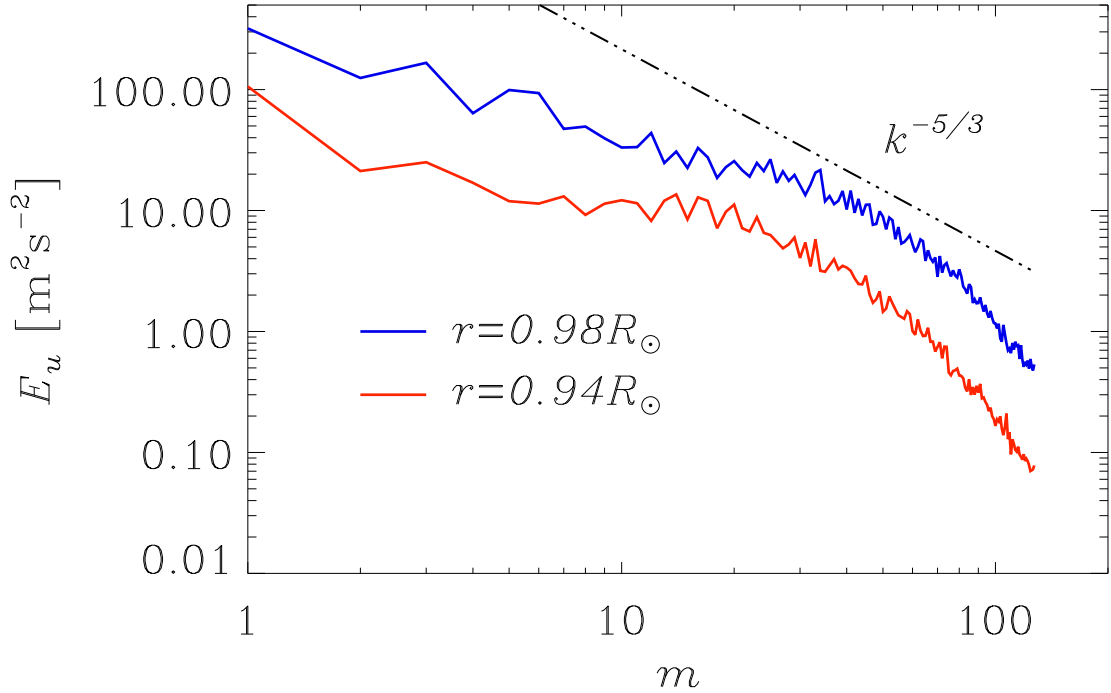}\\
\caption{Power spectrum of the kinetic energy for the model ~N1. The blue and 
red lines correspond to $r=0.98\Rs$ and $r=0.94\Rs$, respectively. 
The dotted-dashed line indicates the Kolmogorov $5/3$ turbulent decay law.}  
\label{fig.pe_nssl}
\end{figure}

The convection pattern, illustrated by the radial velocity field, at 
the top of the domain ($r=0.98\Rs$, upper panel of Fig. \ref{fig.urns}) 
resembles the model ~M1 (middle panel of Fig. \ref{fig.ur}).  
Elongated convective structures are observed at lower latitudes, however, 
these patterns are not the banana cells observed in the model ~M1 but rather
smaller convection cells organized along the banana cells located below the
solar surface.
A snapshot of the radial velocity at $r=0.94\Rs$ clearly shows that the
banana cells still exist below the surface (bottom panel of Fig. \ref{fig.urns}). 
The power spectrum at two different depths for this model shown in 
Fig. \ref{fig.pe_nssl} indicates an excess of power at $m=3$ and $m=5$ 
for $r=0.98\Rs$. There are several peaks of power between 
$10 \lesssim m \lesssim 20$ corresponding to the smallest resolved scales. 
For $r=0.94\Rs$ there is no peak which could be identified with the
banana cells, however the peaks at large scales do not appear. The 
peaks of the smaller scales remain at the same values of the angular degree.

Like in the Sun, the rotation profile shows acceleration at the equator. 
Above $r=0.96\Rs$ the rotation decreases with radius at lower
latitudes (see blue line in Fig. \ref{fig.rpnssl}).  This negative
shear does not remain for latitudes above $15^{\circ}$, and the
contrast of differential rotation in this case is $\chi_{\Omega}=0.087$.
Our interpretation is that the near-surface shear layer is associated
with an inwards transfer of the angular momentum as a consequence of
fast convective motions near the surface in the buoyancy dominated 
regime. The fact that this negative shear
does not spread to high latitudes could be related to the width of
the upper layer. A thicker near-surface layer could induce 
meridional motions transporting the angular momentum in the latitudinal
direction.  It might also be associated with the vertical 
contours of rotation obtained in our model 
(different from the solar rotation profile) or with the 
boundary conditions.  

\section{Discussion}
\label{sec.conc}
Large scale flows in rotating turbulent convection,
like in giant planets and stars, are developed due to collective 
turbulent  effects. These flows are important in astrophysics since they 
could explain the observed patterns of surface differential 
rotation and, in addition, because they could be responsible for the
generation of large-scale magnetic fields via the dynamo process.

In this paper we have explored the development of these large
scale flows in global simulations of a convective envelope whose
stratification resembles the solar interior.  Although our main goal
is study the pattern of differential rotation observed in the Sun,
our results with different rotation rates are applicable also to 
solar-like stars.  

The model considered here has proven to work well in simulating 
stellar interiors. The inviscid numerical method 
allows higher turbulent levels even
with a coarse resolution. Besides, the energy formulation
allows convection development in an slightly super-adiabatic state,
which is expected to happen in an environment where the radiative
flux is small.  

In our numerical experiments we have obtained convective patterns,  
rotation and meridional circulation profiles for solar-like stars
for a broad range of rotation rates. 
The basic properties of the flow structure are compatible with previous
results of rotating convection obtained with different numerical codes
\citep[e.g.,][]{KMGBC11,GWA13}.  The Reynolds stress components
computed from our simulations also agree with previous results
in spherical geometry \citep[e.g.,][]{KMGBC11} and partially 
agree with the mean-field theory of Lambda-effect \citep{RK07}.

We have explained the develpment of differential rotation as
a consequence of the competition between buoyancy and Coriolis 
forces. This balance results in different Reynolds stress 
profiles which define the angular momentum distribution.  For
slow rotating models where buoyancy dominates, the most important
contribution to the angular momentum flux comes from the meridional
circulation, which results from the turbulent correlation 
$\brac{u_r'u_{\theta}'}$.
For large Rossby numbers the results show either one (model ~L1) or
two (models ~L2 and ~L3) meridional cells per hemisphere.

For the rapidly rotating models the correlations $\brac{u_{\phi}'u_{\theta}'}$
and $\brac{u_{\phi}'u_r}$ dominate, and  $\brac{u_r'u_{\theta}'}$ might
have only a marginal contribution. The differential rotation forms due to
these components of the Reynolds stress and the meridional flow
is formed due to inertial forces due to the rotation \citep{MH11}. 
The meridional circulation exhibit several cells forming
at each hemisphere. This result was found in other global convection 
simulations  \citep[e.g.,][]{BT02,KMGBC11,GCS10}. However, the formation
of several circulation cells is at odds with hydrodynamic mean-field 
models.

In order to obtain a solar-like differential rotation profile,
our simulations require convective velocities of the order of 
m/s in the bulk of the convection zone. The Rossby number associated
with this rotation rate and $\urms$ is of the order of $10^{-2}$ 
which defines the Sun as a fast rotator, see Fig. \ref{fig.rovsro}.
Furthermore, the models close to the transition between anti-solar
and solar-like differential rotation also show multiple circulation
cells. Recent observational results also indicate that
the Sun has more than one cell \citep{Ha12,ZBKD12,ZBKD13}. Further studies 
are required, however, to clarify this issue.

We have also shown that the transition between anti-solar to solar-like
rotation could also be obtained by changing the forcing of the 
system, i.e., making convection more or less vigorous (models 
~L3, ~L8 and ~L9). 
However, as mentioned before, for the solar
rotation rate, only the ambient state defined in Eqs. (\ref{eq.ps} to 
\ref{eq.the}) results in a fair agreement with the observed 
properties described in Introduction as items (i) to (iv).
In the model ~L6 we have been able to satisfy the properties
(i) and (iii). An important result is the formation of the
tachocline at the interface between stable and unstable layers.

In all the models presented here the rotation contours have cylindrical
shape. Although at just below the tachocline a latitudinal gradient
in $\Theta'$ is observed in the rapid rotating simulations, 
it is not transported to the inner 
convection zone, thus being insufficient to break the columnar rotation.
The solar rotation property (ii) is not reproduced for our model.
A different choice of the ambient state $\Theta_e$ has proven to 
be able to create radial contours of rotation at middle an high latitudes
\citep{RCGS11}. On the other hand, \cite{WKMB13} have obtained conical
rotation profiles at lower latitudes by considering an extended 
domain where the solar convection zone is surrounded by a simplified
coronal model.  We believe that a more fundamental property
(including the contribution of the magnetic field) of the plasma
could be the responsible for the radial rotation profiles.  A detailed 
study of this will be the subject of a forthcoming paper. 

To tackle the formation of the near-surface shear layer, property 
(iv), we have constructed a model with an extended radial domain up
to $r=0.985\Rs$. In the outer 6\% of the domain the entropy of the
ambient state decreases rapidly with radius. The results of the model ~N1 show
that the Rossby number increases in this thin layer to values
above $\Ro=0.1$. This indicates that convection is vigorous
in this region,  evolves on a short timescale, and is only 
marginally affected 
by rotation. In this case, above $r=0.96\Rs$ the angular velocity
decreases with radius at near equatorial
latitudes $\lesssim 15^{\circ}$. Due to the thin width of the
layer no poleward meridional circulation (like the one observed in slow
rotating models) is formed due to this negative
shear.  Similar results have been obtained by \cite{KMB11} and 
\cite{GW12} with the use of a strong density stratification. 
Although we keep the adiabatic background stratification constant
in the energy equation, the density constrast associated with
the ambient state, $\rho_e$, increases. 
Thus, both methods seem to be equivalent. 
In order to reproduce a near-surface shear layer
spaning over all latitudes a further extended radial domain is
perhaps required. Tilted contours of rotation in the bulk of the
convection zone might also facilitate the formation of this
layer such as it is observed.

We have tested the convergence of our 
models by increasing the resolution of the model L6 by a factor
of $2$ and $4$ in models ~M1 and ~H1, respectively.  The results 
show that for higher resolutions the solar-like differential rotation profile 
disappears in the deep convection zone and is reduced at the surface. 
We noticed that that meanwhile the Rossby number
remains roughly constant, the contribution of the Reynolds 
stress flux decreases and the meridional circulation flux
increases.  This indicates that the smaller scales resolved when
the turbulence of the system is increased are modifying the 
mixing of angular momentum in an unexpected way. This issue
has been found and studied in previous works 
\citep{METC00,EMT00,BT02}, by exploring different parameter
regimes and boundary conditions.  \cite{BT02} found that a 
solar-like latitudinal difference in $\Omega$ could be achieved
at higher Reynolds number by decreasing the Prandtl number.
This implies a more homogeneous and efficient exchange of 
heat and filtering certain scales. This kind of analysis 
involves the use of explicit dissipative terms with coefficients
having eddy values (i.e., a different SGS model) and is 
subject for our future work.

\acknowledgments
We thank P. Charbonneau for his valuable comments on this
paper and J-F Cossette for his important
help in the construction of the spherical model. We 
also thank the anonymous referee for his/her valuable comments
that have improved this paper.
GG acknowledges NSF and NORDITA for travel support. 
This work was supported by the NASA grants NNX09AG81 and
NNX09AT36G. All the simulation
here were performed in the NASA cluster Pleiades.

\bibliographystyle{apj}
\bibliography{bib}

\clearpage
\end{document}